\documentclass[a4paper,fleqn,usenatbib]{mnras}

\usepackage[T1]{fontenc}
\usepackage{ae,aecompl}

\usepackage{newtxtext,newtxmath}

\usepackage{graphicx}
\usepackage{subfigure}
\usepackage{arydshln}
\usepackage{gensymb}

\title{Evidence for a TDE origin of the radio transient Cygnus A-2}

\author[M.N. de Vries]{
M. N. de Vries$^{1}$ \thanks{E-mail: m.n.devries@uva.nl},
M. W. Wise$^{2,1}$, 
P. E. J. Nulsen$^{3,4}$,
A. Siemiginowska$^{3}$, 
A. Rowlinson$^{1,5}$ 
\newauthor and C. S. Reynolds$^{6}$
\\
$^{1}$ Astronomical Institute "Anton Pannekoek", University of Amsterdam, Science Park 904, 1098 XH Amsterdam, The Netherlands \\
$^{2}$ SRON Netherlands Institute for Space Research, Sorbonnelaan 2, 3584 CA, Utrecht, The Netherlands \\
$^{3}$ Harvard-Smithsonian Center for Astrophysics, 60 Garden Street, Cambridge, MA 02138 \\
$^{4}$ ICRAR, University of Western Australia, 35 Stirling Hwy, Crawley, WA 6009, Australia \\
$^{5}$  ASTRON, Netherlands Institute for Radio Astronomy, Postbus 2, 7990 AA, Dwingeloo, The Netherlands \\
$^{6}$ Institute of Astronomy, University of Cambridge, Cambridge, CB3 0HA, UK
}

\date{Accepted XXX. Received YYY; in original form ZZZ}

\pubyear{2017}

\begin{document}
\label{firstpage}
\pagerange{\pageref{firstpage}--\pageref{lastpage}}
\maketitle

\begin{abstract}
In 2015, a radio transient named Cygnus A-2 was discovered in Cygnus A with the Very Large Array. Because of its radio brightness ($\nu F_{\nu} \approx 6 \times 10^{39}$ erg s$^{-1}$), this transient likely represents a secondary black hole in orbit around the AGN. Using {\it Chandra} ACIS observations from 2015 to 2017, we have looked for an X-ray counterpart to Cygnus A-2. The separation of 0.42 arcsec means that Cygnus A-2 can not be spatially resolved, but by comparing the data with simulated \texttt{marx} data, we put an upper limit to the 2-10 keV X-ray luminosity of Cygnus A-2 of $1 \times 10^{43}$ erg s$^{-1}$. Using the Fundamental Plane for accreting black holes, we find that our upper limit to the X-ray flux of Cygnus A-2 in 2015-2017 disfavours the interpretation of Cygnus A-2 as a steadily accreting black hole. We suggest instead that Cygnus A-2 is the radio afterglow of a tidal disruption event (TDE), and that a peak in the 2-10 keV luminosity of the nuclear region in 2013, when it was observed by {\it Swift} and {\it NuSTAR}, is X-ray emission from the TDE. A TDE could naturally explain the X-ray light curve of the nuclear region, as well as the appearance of a short-lived, fast, and ionized outflow previously detected in the 2013 {\it NuSTAR} spectrum. Both the radio and X-ray luminosities fall in between typical luminosities for 'thermal' and 'jetted' TDE types, suggesting that Cygnus A-2 would be unlike previously seen TDE's.
\end{abstract}


\section{Introduction}
The central engine of the radio-bright galaxy Cygnus A is generally understood to be a heavily obscured, broad-line Active Galactic Nucleus (AGN) \citep{Antonucci1994, Ueno1994, Ogle1997}. Because of the heavy obscuration by the giant elliptical host galaxy, the AGN is not directly visible at optical, UV and soft X-ray wavelengths \citep{Young2002, Canalizo2003}. In contrast to the extreme radio power of the galaxy, the AGN itself is only moderately luminous with a bolometric luminosity $L_{\rm Bol} = 10^{45}$ erg s$^{-1}$ \citep{Privon2012}.

In July 2015, \cite{Perley2017} discovered a transient point source in Cygnus A with the Karl G. Jansky Very Large Array (VLA). The transient, named Cygnus A-2, is located 0.42 arcsec (460 pc) southwest of the primary AGN. Follow-up observations in 2016 with VLA and the Very Long Baseline Array (VLBA) confirmed the presence of the source, and the flux and spectral shape showed no obvious variability within that year. The transient was not observed in previous VLA observations, most recently in 1997, and therefore appeared sometime between 1997 and 2015. 

The radio luminosity $\nu F_{\nu} \approx 6 \times 10^{39}$ erg s$^{-1}$ makes Cygnus A-2 a rather bright radio transient. The high luminosity puts a strong constraint on the origin of Cygnus A-2. \cite{Perley2017} investigated whether Cygnus A-2 could be the radio afterglow of a supernova, although a supernova with such a high radio luminosity would be a rare and unlikely event. The favoured explanation is that Cygnus A-2 is a secondary black hole, orbiting the primary AGN. The black hole could have brightened due to a steady increase in accreting material in the past 20 years, or through the sudden disruption of a star by the black hole, known as a tidal disruption event (TDE). 

A point source coinciding with the location of Cygnus A-2 was observed in the infrared with the \textit{Keck} II AO system in 2003, in the $J$, $H$ and $K'$  bands \citep{Canalizo2003}. The authors determined that this same point source is also present in {\it Hubble Space Telescope} (HST) observations from 1996 and 1997. Based on the SED and the total luminosity, the most likely interpretation for this point source is that it is a tidally stripped galaxy core, that is in the process of merging with the giant elliptical galaxy. This supports the secondary black hole hypothesis put forth by \cite{Perley2017}. Alternatively, the point source as seen by \textit{Keck} and HST could represent the light from the secondary black hole itself.

Additionally, between 1 and 3 arcsec northwest and southeast of the AGN, a bipolar region with narrow emission lines has been observed in optical observations with the HSTS \citep{Tadhunter1994, Jackson1998}. The regions have an approximately paraboloidical shape and are oriented around the jet axis, suggesting that they scatter the light of the obscured AGN into our line of sight.

The nuclear region of Cygnus A has been studied at X-ray wavelengths as well, although these observations are more limited in their angular resolution. In \textit{Chandra} ACIS observations, the AGN is moderately piled up with the standard 3.2s frame time. One 0.4s frame time observation from May 2000 has previously been used by \cite{Young2002} to study the region. At energies below 2 keV, the same bipolar emission regions are observed as with HST, while at energies above 2 keV, the AGN itself becomes visible. 

In 2013, the {\it Nuclear Spectroscopic Telescope Array} \citep[{\it NuSTAR};][]{Harrison2013}, provided a new view of the AGN of Cygnus A at energies up to 70 keV. \cite{Reynolds2015} have compared this observation with an \textit{XMM-Newton} observation from 2005. The two observations differ in two notable ways. Firstly, the 2-10 keV X-ray luminosity appears to have doubled between 2005 and 2013. Secondly, subtle redshifted emission and blueshifted absorption features were detected around the iron K$\alpha$ line at 6 keV in the \textit{NuSTAR} spectrum. These features are well fit by a fast, highly ionized wind with a high column density. This wind would presumably have started up in the 7 years between the two observations. We have summarized the observed properties of the X-ray emission of the AGN as measured by \textit{Chandra}, \textit{XMM-Newton}, and \textit{NuSTAR} in Table \ref{AGN:tab:lumin}. 

\begin{table}
\caption{Overview of previously measured X-ray properties of the AGN, measured by 3 different X-ray instruments. }
\label{AGN:tab:lumin}
\begin{tabular}{c c c c c c}
\hline \hline
Date & Ref & Instrument & $N_{H}$ & $L_{2-10 \rm keV}$ & $\Gamma$ \\ 
& & & $10^{23}$ cm$^{-2}$ & erg s$^{-1}$ &\\ \hline
2000 05 & 1 &  {\it Chandra} ACIS & $ 2.0_{-0.1}^{+0.2}$ & $1.9 \times 10^{44}$  & $1.52_{-0.12}^{+0.12}$ \\
2005 10 & 2 & {\it XMM-Newton} & $3.4_{-0.3}^{+0.3}$ & $2.0 \times 10^{44}$ & $1.43_{-0.11}^{+0.11}$ \\
2013 02 & 2 & {\it NuSTAR} & $1.7_{-0.1}^{+0.1}$ & $4.1 \times 10^{44}$ & $1.47_{-0.06}^{+0.13}$ \\ \hline
\end{tabular} \\
\texttt{References} - 1) \cite{Young2002}, 2) \cite{Reynolds2015}

\end{table}

In this paper, we have searched for evidence of X-ray emission from Cygnus A-2, with the aim of shedding more light on the nature of this transient. Of particular interest are \textit{Chandra} ACIS observations from 2000, 2005 and 2015 to 2017: firstly because its angular resolution of 0.5 arcsec is only slightly larger than the 0.42 arcsec separation between the AGN and Cygnus A-2, and secondly because the 2015-2017 observations are from the same epoch as the original discovery of Cygnus A-2 with the VLA. We compare our results with {\it XMM-Newton} and {\it NuSTAR} observations, described in \cite{Reynolds2015}, as well as archival observations from the the X-ray Telescope (XRT) on board the Neil Gehrels Swift Observatory \citep[hereafter referred to as {\it Swift},][]{Gehrels2004, Burrows2005}.

We give an overview of the \textit{Chandra} observations in section \ref{AGN:sec:obs}. In section \ref{AGN:sec:impsf}, we image the soft X-ray emission and hard X-ray point source, compare the {\it Chandra} image data with \texttt{marx} simulations, and use the image and the simulated data to look for an X-ray counterpart to Cygnus A-2. In section \ref{AGN:sec:spectral}, we perform an X-ray spectral analysis of the AGN with the \textit{Chandra} data, including modeling the pileup. We compare the 2-10 keV luminosities with \textit{Swift} XRT, \textit{XMM-Newton}, \textit{NuSTAR}, \textit{EXOSAT}, and \textit{ASCA} observations to construct a light curve of the AGN between 2000 and 2017. We also look for signatures of the fast, ionized wind in the \textit{Chandra} data. We discuss the results of the image and spectral analysis in section \ref{AGN:sec:disc}, and conclude in section \ref{AGN:sec:conc}.

Throughout this paper, we have adopted a standard cosmology with $H_{0}$= 69.3 km s$^{-1}$ Mpc$^{-1}$, $\Omega_{M}$ = 0.288, and $\Omega_{\Lambda} = 0.712 $ \citep{Hinshaw2013}. We use a redshift value of z=0.0561 \citep{Stockton1994}. Using this cosmology, the linear scale is 66 kpc per arcminute and the luminosity distance $D_L = 253.2 $ Mpc for Cygnus A.  To facilitate comparison, X-ray luminosities cited from previous work, have been re-calculated for this luminosity distance.

\section{Observations}

\label{AGN:sec:obs}

\begin{table}
\caption{Overview of {\it Chandra} ACIS observations with the AGN close to the aimpoint. ObsIDs 360 and 1707 were taken with the ACIS-S array, all others with ACIS-I. The spectra and images of ObsIDs between dashed lines are considered to be within the same time period and are analysed together. ObsID 1707 is marked with an asterisk to indicate that it is the only short (0.4s) frame time observation.}
\begin{tabular}{ c c c c}
\hline \hline
Date                & Observation ID & Exposure \\ 
(UTC)               &                & (ks)      \\
\hline
2000-05-21T03:13 & 360   & 34.7 \\
2000-05-26T12:28 & 1707* & 9.2  \\
\hdashline
2005-02-15T15:26 & 6225  & 24.3 \\
2005-02-16T13:01 & 5831  & 50.8 \\
2005-02-19T05:10 & 6226  & 23.7 \\
2005-02-21T03:06 & 6250  & 7.0  \\
2005-02-22T11:59 & 5830  & 23.2 \\
2005-02-23T11:52 & 6229  & 22.8 \\
2005-02-25T04:27 & 6228  & 16.0 \\
2005-09-07T04:48 & 6252  & 29.7 \\ 
\hdashline
2015-10-28T11:49 & 17508 & 14.9 \\
2015-11-01T16:48 & 18688 & 34.6 \\
2016-06-13T21:02 & 18871 & 21.8 \\
2016-06-18T17:52 & 17133 & 30.2 \\
2016-06-26T16:13 & 17510 & 37.3 \\
2016-07-10T22:52 & 17509 & 51.2 \\
2016-08-15T22:10 & 17513 & 49.1  \\
2016-09-15T05:46 & 17512 & 66.9 \\ 
2016-11-12T12:56 & 17507 & 32.4 \\
2016-12-13T10:56 & 17514 & 49.4 \\ 
2017-01-20T18:09 & 17135 & 19.8 \\ 
2017-01-26T00:42 & 17136 & 22.2 \\ 
2017-01-28T14:24 & 19996 & 28.6 \\
2017-02-12T05:40 & 19989 & 41.5 \\
2017-05-10T02:01 & 17511 & 15.9 \\
2017-05-13T21:25 & 20077 & 27.7 \\
2017-05-20T23:24 & 17134 & 29.4 \\ 
2017-05-21T17:12 & 20079 & 23.8 \\
\hline
                 & Total & 837  \\
\hline
\end{tabular}
\label{tab:ChandraObs}
\end{table}

More than 2.2 Ms of \textit{Chandra} ACIS data is currently publicly available on the {\it Chandra} Data Archive. However, a significant fraction of this data is pointed at some of the extended features in the system, such as the eastern and western hot spots, the merger region, and the northwestern subcluster CygNW. 

To minimize the size of the PSF, we have selected only those observations pointed at the AGN. This leaves us with a total of 873ks of data. Of those 873ks, 43.9ks were taken in May 2000, the first year of \textit{Chandra} operations. ObsID 1707 is of particular note, because it is the only one with a short frame time of 0.4s, specifically chosen to avoid pileup in the AGN. A second set of observations were taken in 2005, totaling 197.5ks. Finally, a large set of observations of were taken between October 2015 and May 2017, totaling 595.6ks. We show a list of all the observations in Table \ref{tab:ChandraObs}.

Each of these datasets has been reprocessed using \texttt{CIAO} 4.9 and \texttt{CALDB} 4.7.4 \citep{Fruscione2006}. \textit{Chandra's} pointing inaccuracy means that there can be an offset of up to an arcsec from the real coordinates. Therefore, an astrometric correction is necessary to compare the same region between Chandra datasets. We have followed procedure of \cite{Snios2018}, briefly outlined here. ObsID 5831 was chosen as the reference observation, because of the long exposure time and the high number of events. We calculated the coordinates of the center of mass within a 5'' x 5'' square, centered on the AGN. We then reprojected these coordinates to the astrometric position of the nucleus, $\alpha_{\rm nuc}=$ 19:59:28.35648, $\delta_{\rm nuc}= $ +40:44:02.0963 \citep{Gordon2016} with \textit{wcs\_update}. We then cut out a 160'' x 120'' rectangular region of the central AGN of CygA.  The event lists of the other ObsIDs were reprojected onto the sky frame of ObsID 5831. A 0.5 - 7.0 keV image of the reprojected events was then cross-correlated with that of ObsID 5831, and then fitted with a Lorentzian function, to determine the coordinate offset. This coordinate shift was then applied to the event list and the aspect solution file with \textit{wcs\_update}.
	
After the astrometric correction, we applied the following CIAO processing tools. For each ObsID, a new badpix file was built with \textit{acis\_build\_badpix}. We applied this badpix file, and the latest gain and CTI corrections with \textit{acis\_process\_events}. We created a new level 2 event file by filtering for good grades (0,2,3,4,6). After that, we filtered for GTIs with the \textit{deflare} procedure.

\section{Source imaging and the PSF}
\label{AGN:sec:impsf}
\subsection{Imaging}

\label{AGN:subsec:imaging}

\begin{figure*}
   \includegraphics[width=0.75\textwidth]{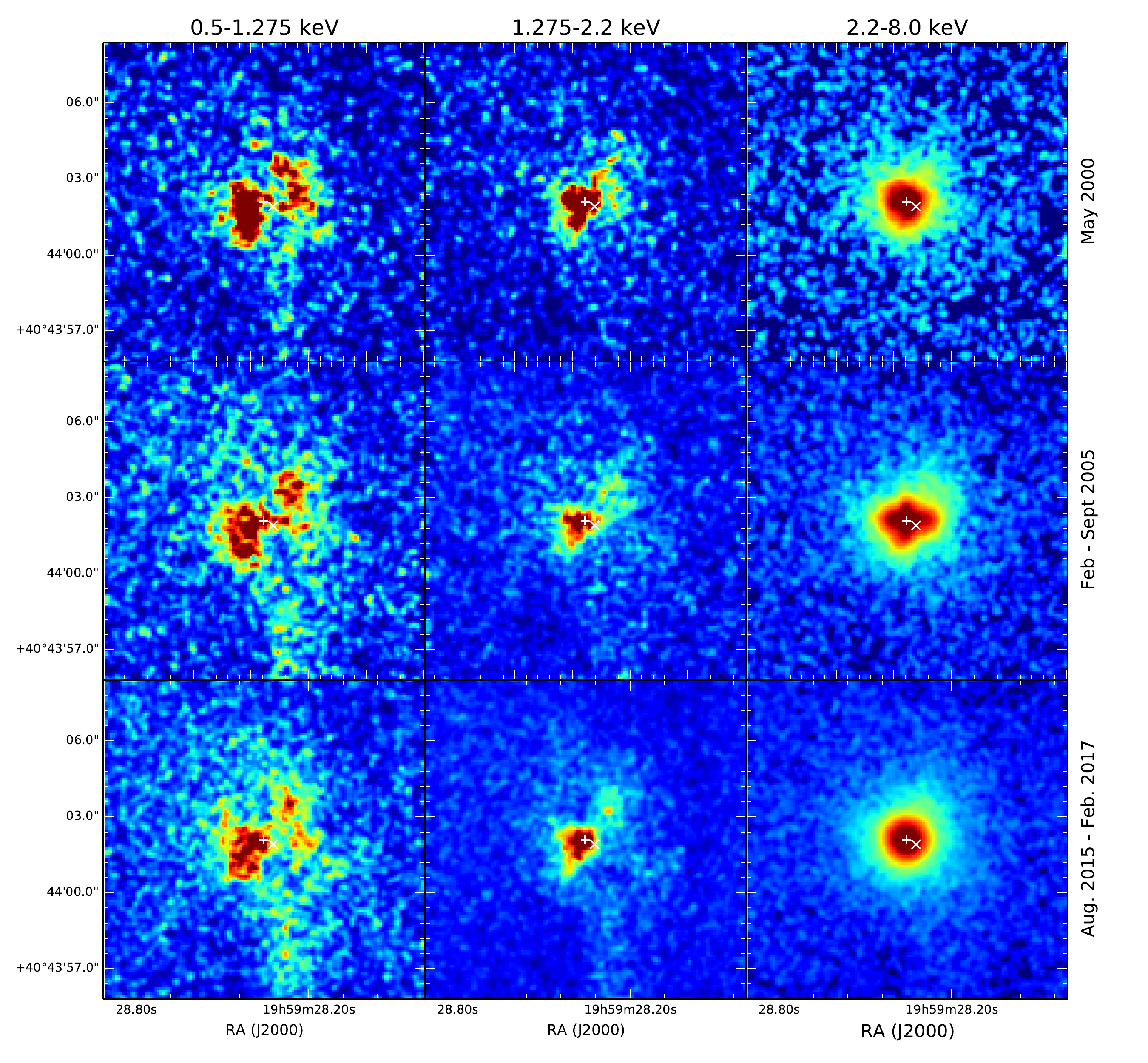}

  \caption{12.5" x 12.5" images showing the central region flux over time. The image has been binned at 0.2 ACIS pixel size in both dimensions, and smoothed by a  0.5'' FWHM Gaussian. The astrometric positions of the AGN and Cygnus A-2 are respectively indicated by a cross and an X.} 
  
    \label{fig:rgb_images}

\end{figure*}

We imaged the central region of Cygnus A in 3 different energy bands: the soft (0.5-1.275 keV), intermediate (1.275-2.2 keV) and hard (2.2-8.0 keV) bands, chosen after \cite{Young2002}. The ACIS blank-sky backgrounds were used as backgrounds for these images. These backgrounds are imported from CALDB with the \textit{acis\_bkgrnd\_lookup}, and  scaled to the event files by using the counts between 10.0 - 12.0 keV. The event files and backgrounds were then reprojected to the common tangent point of ObsID 360.

For each ObsID, we extracted a spectrum from a region with a radius of 6 arcsec, centered on the source. We then fit a model to them to obtain energy weightings for the exposure maps. These fits are not meant to find the physical characteristics of the source, but only to find a model curve that approximates the spectral shape, to obtain a rough energy weighting for the exposure-corrrected images. To the 0.5-2.2 keV band, we fit a power law multiplied by the galactic absorption component, which we froze at $3.1 \times 10^{21}$ $N_{\rm H}$ cm$^{-2}$ \citep{Dickey1990}. To the 2.2-8.0 keV band, we fit the same model, but we leave the absorption component as a free parameter, as the hard X-rays come from the highly absorbed central source \citep{Young2002}. 

These model curves were used to obtain spectral weightings, and we used these to create instrument maps and exposure maps with the tools \textit{mkinstmap} and \textit{mkexpmap}. The exposure maps were then reprojected to the tangent point of ObsID 360.

We binned each each event file and blank-sky background into the 3 energy bands, for each of the 3 time bins of 2000, 2005, and 2015-2017.  We then sliced a 12.5 x 12.5'' image of the central region for each ObsID, spatially binned at 0.2 native pixel resolution. We used \textit{dmregrid} to bin the exposure maps to the same resolution. We subtracted the backgrounds from the data and divided by the exposure map to obtain images of the central region. The images are shown in Fig. \ref{fig:rgb_images}.

The nucleus, visible in the 2.2-8.0 keV band, has an irregular, non-pointlike shape in the 2000 and 2005 observations. These asymmetries are the strongest in the 2005 observations, and they show up in all of the the individual ObsIDs from 2005, indicating they are not astrometry related. Further inspection of these observations shows that they were carried out with a large SIM\_Z offset, about 10 mm larger than the nominal value.
At an off-nominal offset, the curved detector focal place intersects the focal surface of the mirrors differently, causing a different defocus. We believe this to be responsible for the irregular PSF shape in the 2000 and 2005 observations.

\subsection{Simulations and the pileup fraction}

\label{AGN:subsec:PSF}

\begin{figure}
\includegraphics[width=0.5\textwidth]{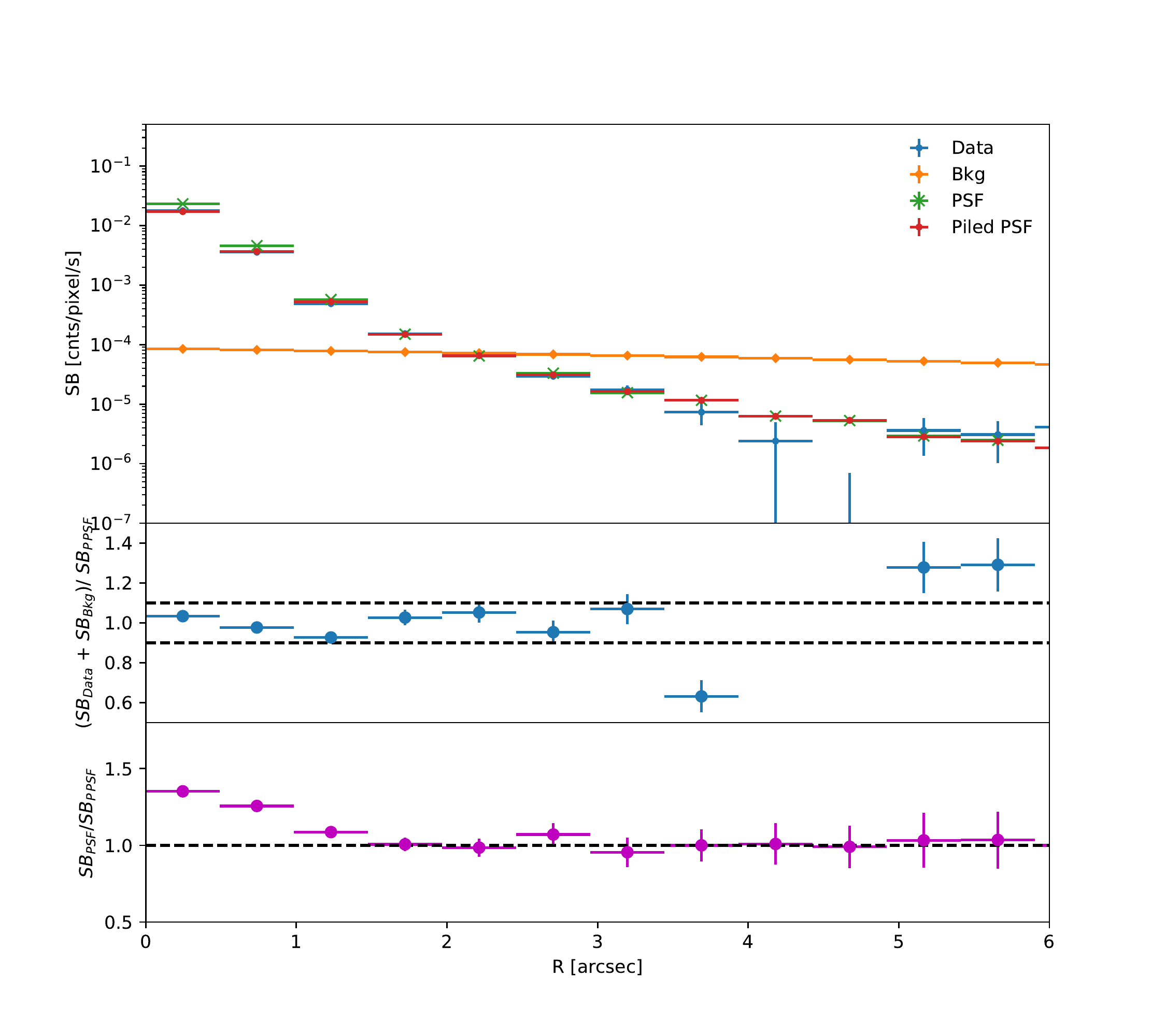}

  \caption{Top: Surface brightness profile of the composite 2015-2017 2.2-8.0 keV data in blue, the background in yellow, and the simulated PSF (green) and piled simulated PSF (red) from the \texttt{marx} simulations. Middle: Residual of the data divided by the piled PSF. Bottom: The ratio  between the unpiled and piled PSF.}
  
   \label{fig:2015_2017_psf}

\end{figure}

We have used the simulation tool \texttt{marx} \citep{Davis2012} to simulate the spatial and spectral distributions of events for each observation. The purpose was to determine the size of the region in which pileup affects the spectrum and to be able to compare the image to data, to look for evidence of any hard X-ray emission that is more extended than the the point source.

As discussed in Section \ref{AGN:subsec:imaging}, the 2000 and 2005 observations show an unusual PSF shape, with wings to the north, south, east and west. We have not been succesful in reproducing this irregular PSF shape with \texttt{marx}. In this section we therefore focus on the 2015-2017 observations, totaling around 595.6 ks of data. As these observations were taken shortly after the discovery of Cygnus A-2, these are the most likely to show any X-ray emission from the transient.

Based on the images of the source, it is apparent that the soft X-rays primarily come from an extended region, while the hard X-rays primarily come from the point source. If pileup was not a factor, the hard X-rays could be simulated as a point source and the soft X-rays could be ignored. However, because of pileup,  both the spatial and spectral distribution of the source emission affect the spectrum extracted from a region. Thus the soft X-ray photons coming from the extended region need to be taken into account when simulating the source. We therefore performed two separate simulations for each ObsID. The soft, extended emission between 0.5 and 2.2 keV was simulated by using the exposure-corrected image of the source ({\tt SourceType=IMAGE}), and an extracted spectrum, both obtained from ObsID 1707 and filtered for counts between 0.5 and 2.2 keV. {\tt Marx} matches the brightness distribution of the simulated source to the input image, so that the simulated source has the same shape, and the same spectral distribution. The 2.2-8.0 keV emission was simulated as a point source ({\tt SourceType=POINT}). The input spectrum was again obtained from ObsID 1707, filtered for counts between 2.2 and 8.0 keV. 

Other parameters relevant for the simulation, like the aspect solution file, the exposure time, detector type, and SIM offsets, were all used in the simulation by reading the header of each event file. The AspectBlur parameter, which blurs the PSF by a fraction of an arcsecond to account for the uncertainty in the determination of the aspect solution, was set to 0.20 arcsec. This is the recommended value for ACIS-I.

After both simulations were performed for a given ObsID, the simulations were combined with the tool \textit{marxcat}. We then applied the pileup tool \textit{marxpileup} to the combined simulation and binned images of the same size as the data, a 12.5 x 12.5'' image at 0.2 native pixel resolution. Finally, we filtered each simulated observation for events between 2.2 and 8.0 keV, and summed all the images of individual observations to obtain the combined 2.2-8.0 keV simulated data of the 2015-2017 observations. Because the source is embedded in the intracluster medium (ICM), there are a significant number of background counts, especially at radii beyond 3 arcsec. Therefore, we modeled the background as a linear function $bkg(r) = a * r + b$ that is fit to the data - model residual. 

The results of our comparison between data and the simulation are shown in Fig. \ref{fig:2015_2017_psf}. We note that, to achieve the best match between the data and the simulation, we iterated over the SourceFlux parameter, which renormalizes the total flux of the model. Because the soft X-ray emission is a few kiloparsec in size, we do not expect that it varied significantly between 2000 and 2015, and so we leave this input spectrum as it is. However, the AGN itself could well have changed in brightness. We iterated several times to find the best match between the piled PSF and the image data, and ultimately found the best match by setting the SourceFlux to about 90\% of that of the input spectrum of ObsID 1707. In section \ref{AGN:subsec:spectral}, we compare this estimate against the luminosities from the spectral analysis.

In the bottom panel of Fig. \ref{fig:2015_2017_psf}, we have compared the PSF and the piled PSF to determine the radius within which pileup is significant, approximately 1.3-1.5 arcsec. Furthermore, we determine that the piled PSF contains 78\% of the counts of the unpiled PSF, yielding a pileup fraction of 22\%. This provides us with an estimate of the pileup fraction to compare against in section \ref{AGN:subsec:spectral}. 

\subsection{Searching for an X-ray counterpart to Cygnus A-2}

\label{AGN:subsec:cygA2}
\begin{figure}
   \includegraphics[width=0.5\textwidth]{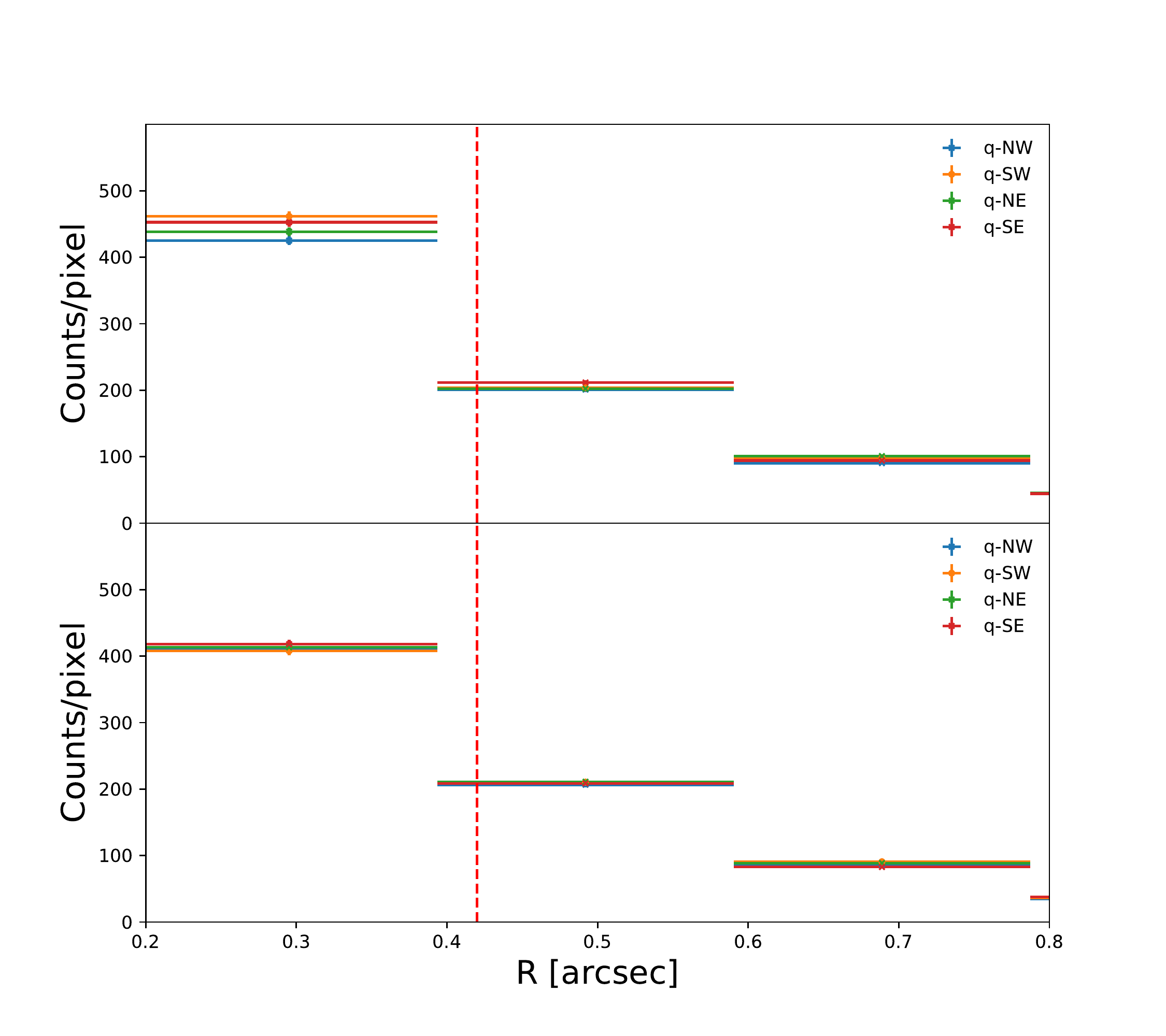}

  \caption{Top: Counts per pixel profiles of the 2015-2017 data, divided in NW, NE, SW, and SE quadrants. The dashed red line indicates the location of Cygnus A-2, at 0.42 arcsec, in the SW quarter. Bottom: as above, but for the image simulated with \texttt{marx}.}
  
    \label{fig:sb_quadrants}

\end{figure}

\begin{figure*}
\centering
   \includegraphics[width=1.1\textwidth]{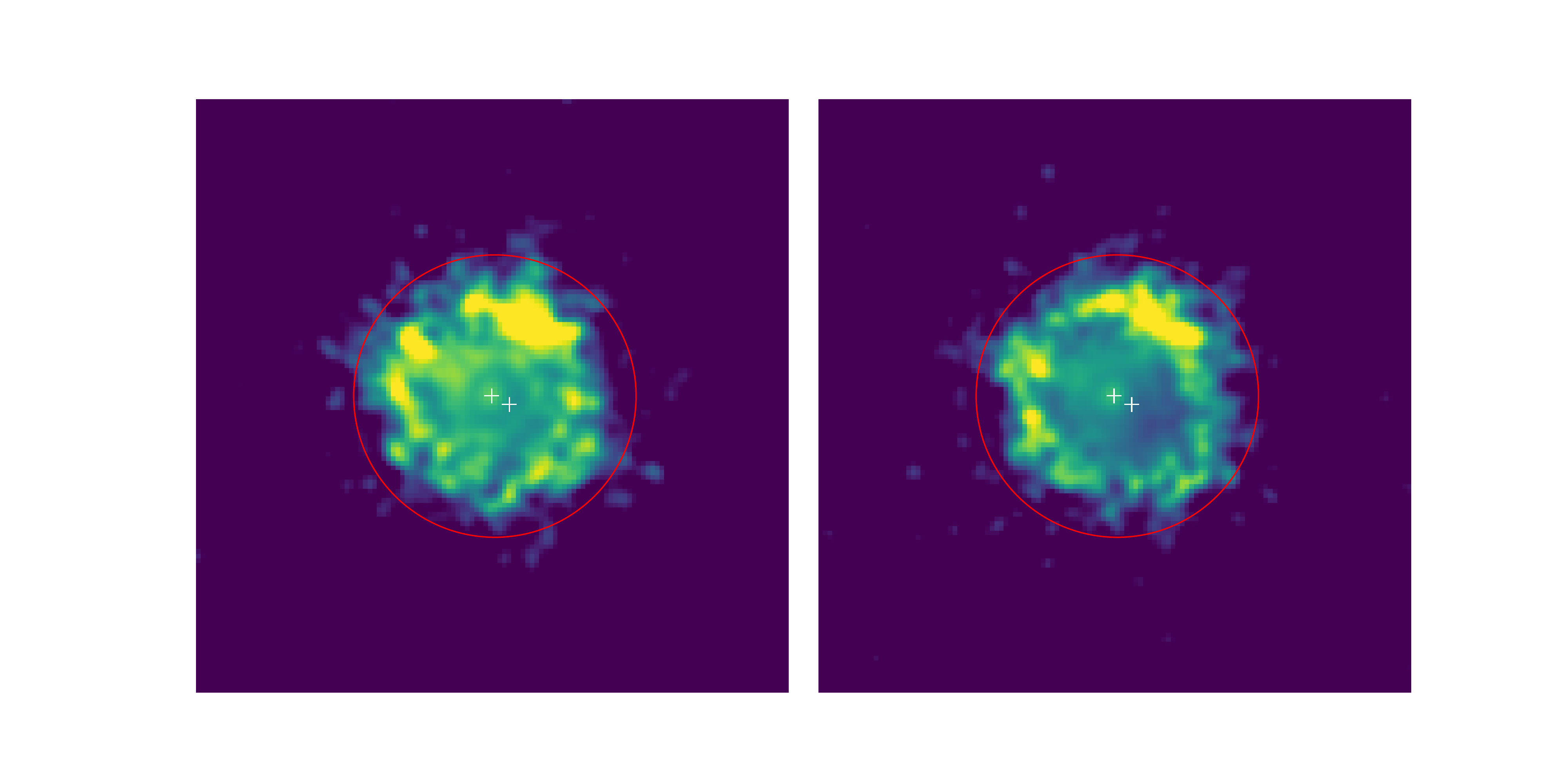}

  \caption{Left: residual of the 2.2-8.0 keV image divided by the simulated image.  Right: residual of the 2.2-8.0 keV image divided by the simulated image with an additional point source, with 5\% the flux of the AGN point source, at the location of Cygnus A-2. Both images have been smoothed by a 0.5 arcsec Gaussian. The white crosses indicate the astrometric locations of Cyg A and Cyg A-2, the red circle indicates a radius of 3 arcsec.}
  
    \label{AGN:fig:psfres}

\end{figure*}

We have used the 2015-2017 data to search for evidence of Cygnus A-2. In the 2000 and 2005 observations, the irregular shape of the PSF makes this comparison more difficult, as we were not able to reproduce the shape in the {\tt Marx} simulations. The 2015-2017 data is the most obvious place to look for Cygnus A-2 in either case, because it is the time period closest to detection. If Cygnus A-2 is emitting X-rays, it would elongate the circular shape of the point source. We therefore look for hints of an elongated point source, by comparing the 2.2-8.0 keV image data from section \ref{AGN:subsec:imaging} with the simulated image from section \ref{AGN:subsec:PSF}. 

In the 2015-2017 image, we determined the center coordinate by calculating the center of mass of the image. We then divided the image into quadrants NW, NE, SW, and SE, using the center of mass as the center point. The transient is located at 0.42 arcsec southwest of the center, and would thus be expected to show most clearly in the SW quadrant. We calculated the SB profile for each of the quadrants. We did the same for the simulated and piled PSF in order to understand the order of magnitude of asymmetries in the PSF versus the image. We show the SB profiles of the quadrants of both the image and the PSF in Figure \ref{fig:sb_quadrants}.

While the SE quadrant has the highest surface brightness at 0.42 arcsec, the SW has the same surface brightness to within just a few percent. In the simulated PSF, we see similar kinds of deviation, although slightly smaller than in the data. Around 0.4 arcsec, the differences are of order 1-2\%.  It is likely that at least some of these differences are caused by the fact that the center of the source is not perfectly in the center of a single subpixel, which means the different quadrants are not perfect quarters of the circle. In the range from 0.2 to 0.6 arcsec, the total counts difference between the SW and NE quadrants in the data is about 5\%. We therefore expect the maximum observed 2.2-8.0 keV flux to be not more than 5\% of that of the AGN.

As support for this estimate, we revisited the \texttt{marx} simulations of section \ref{AGN:subsec:PSF}. We created a new simulated PSF by adding a secondary point source at the location of Cygnus A-2 to the PSF simulations, and scaled the input spectrum of that point source to 5\% of the primary point source spectrum. We then created two residual images: the first one by dividing the 2015-2017 2.2-8.0 keV data by the original simulated PSF, and the second one by dividing the data by the simulated PSF with the transient added. The left and right panels of Fig. \ref{AGN:fig:psfres} shows these two residual images. The left panel shows a similar structure to the surface brightness profiles in Fig. \ref{fig:2015_2017_psf}: a decent match up to a radius of 2.5-3 arcsec. At larger radii, the background dominates. The residual appears to be roughly radially symmetric. The right panel of Fig. \ref{fig:2015_2017_psf} shows the data divided by the PSF with the transient added. In this residual, there is a visible deficit at the location of the secondary point source, particularly noticeable towards the southwest. We conclude that a source with a flux more than 5\% would have been visible in the comparison between the data and the PSF.

\section{Spectral analysis}
\label{AGN:sec:spectral}

\subsection{Chandra observations}
\label{AGN:subsec:spectral}
\begin{figure*}
\centering     
\subfigure[2000 SF]{\label{fig:a}\includegraphics[width=88mm]{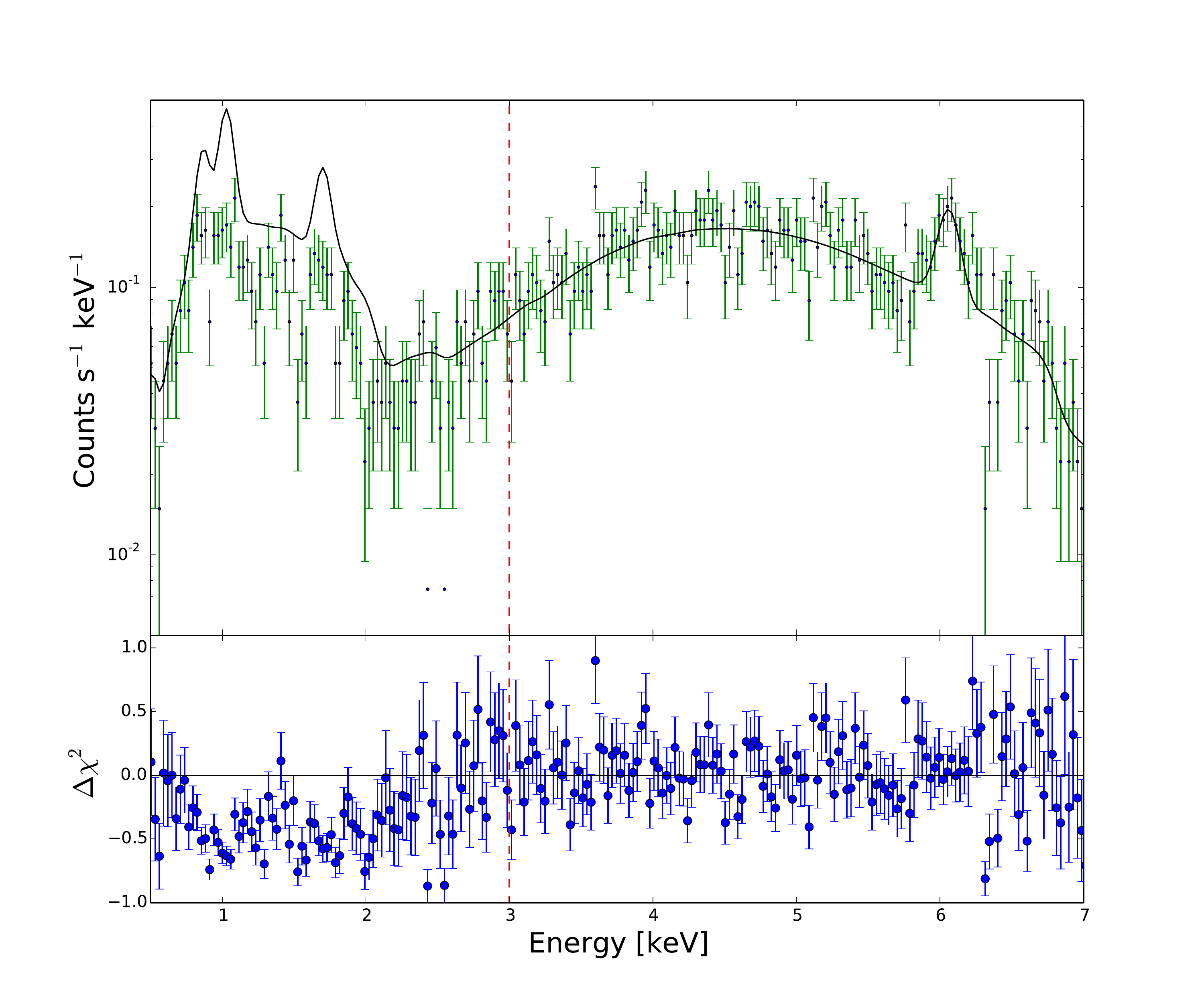}}
\subfigure[2000 NF]{\label{fig:b}\includegraphics[width=88mm]{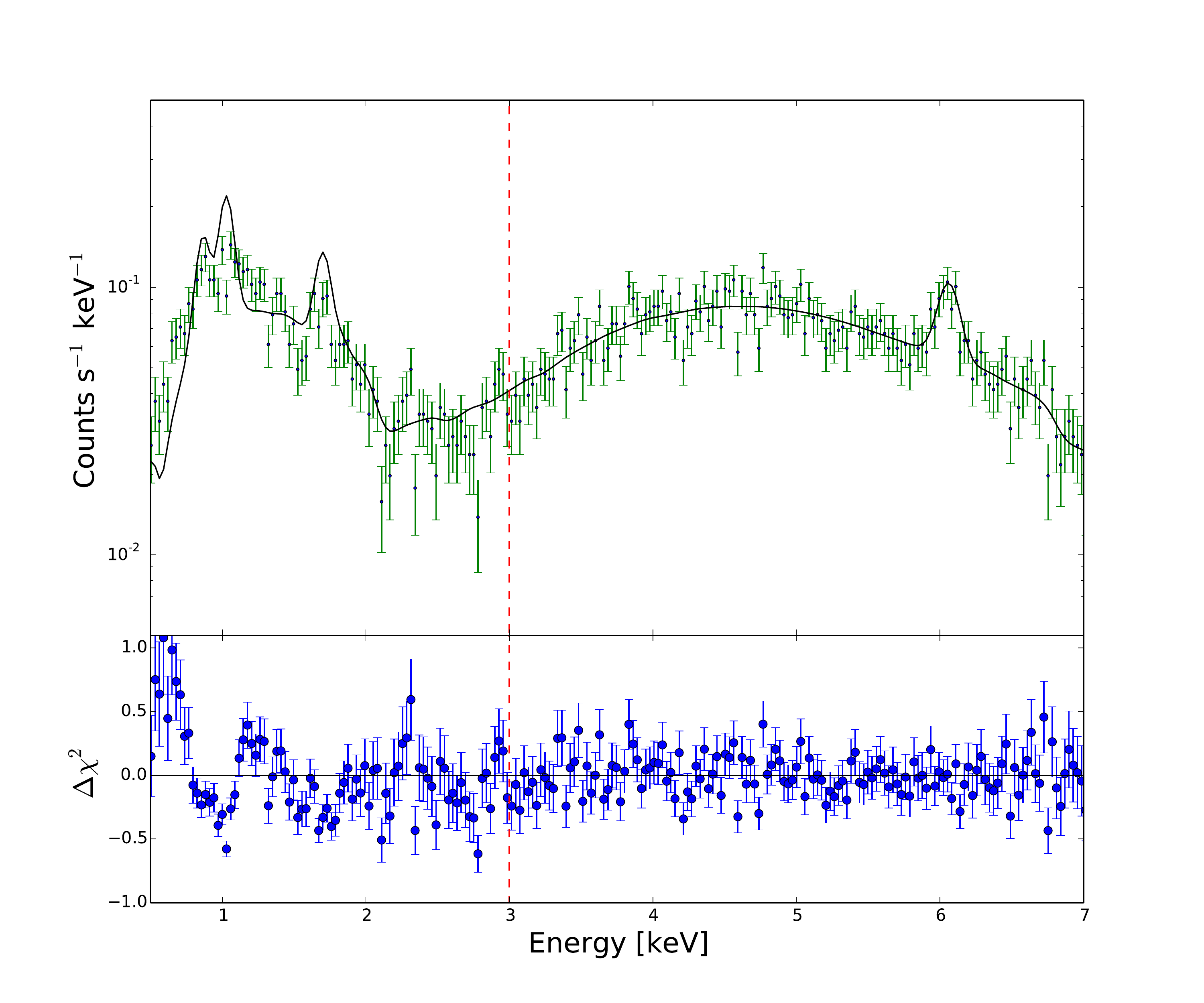}}
\subfigure[2005]
{\label{fig:c}\includegraphics[width=88mm]{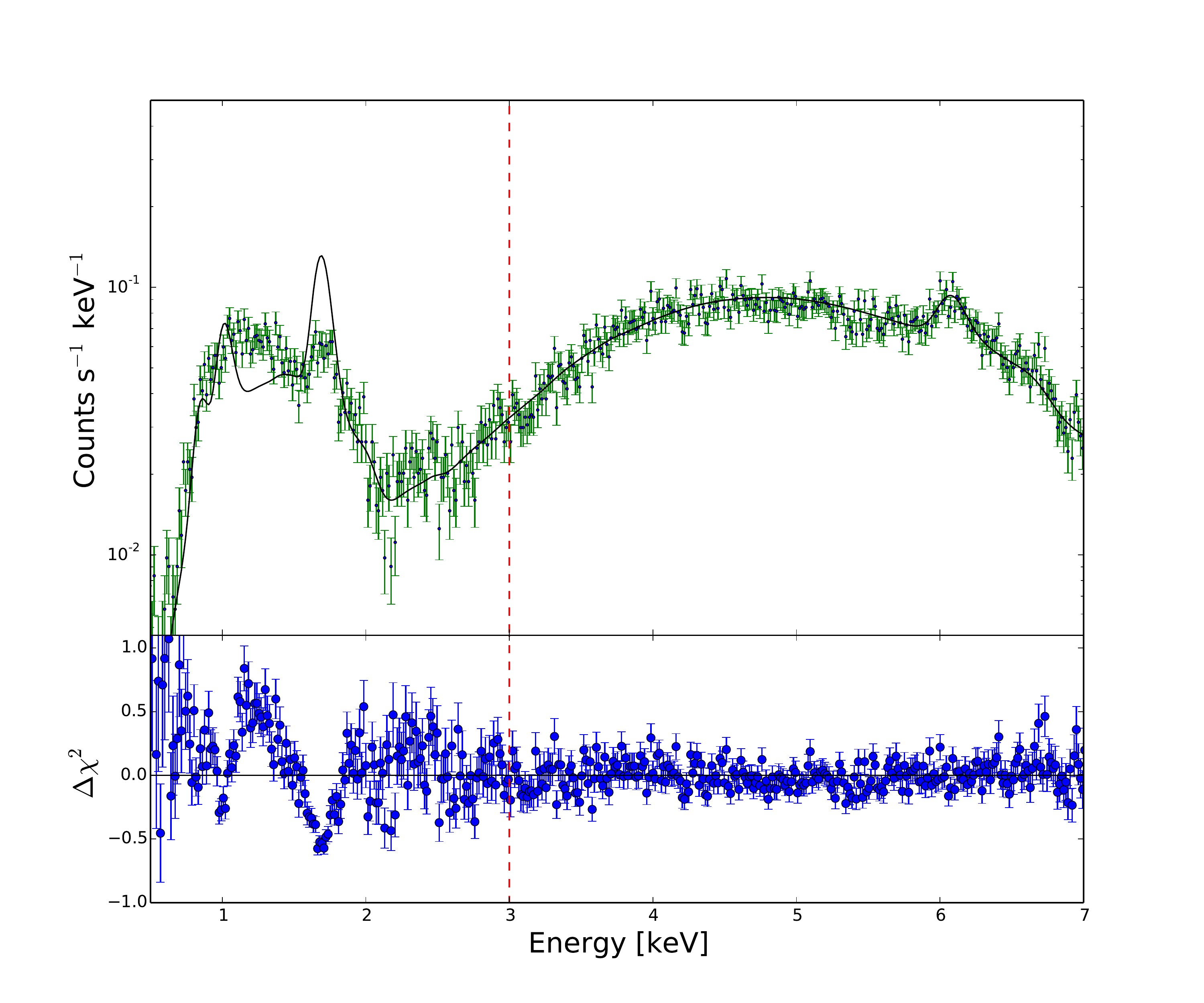}}
\subfigure[2015-2017]
{\label{fig:d}\includegraphics[width=88mm]{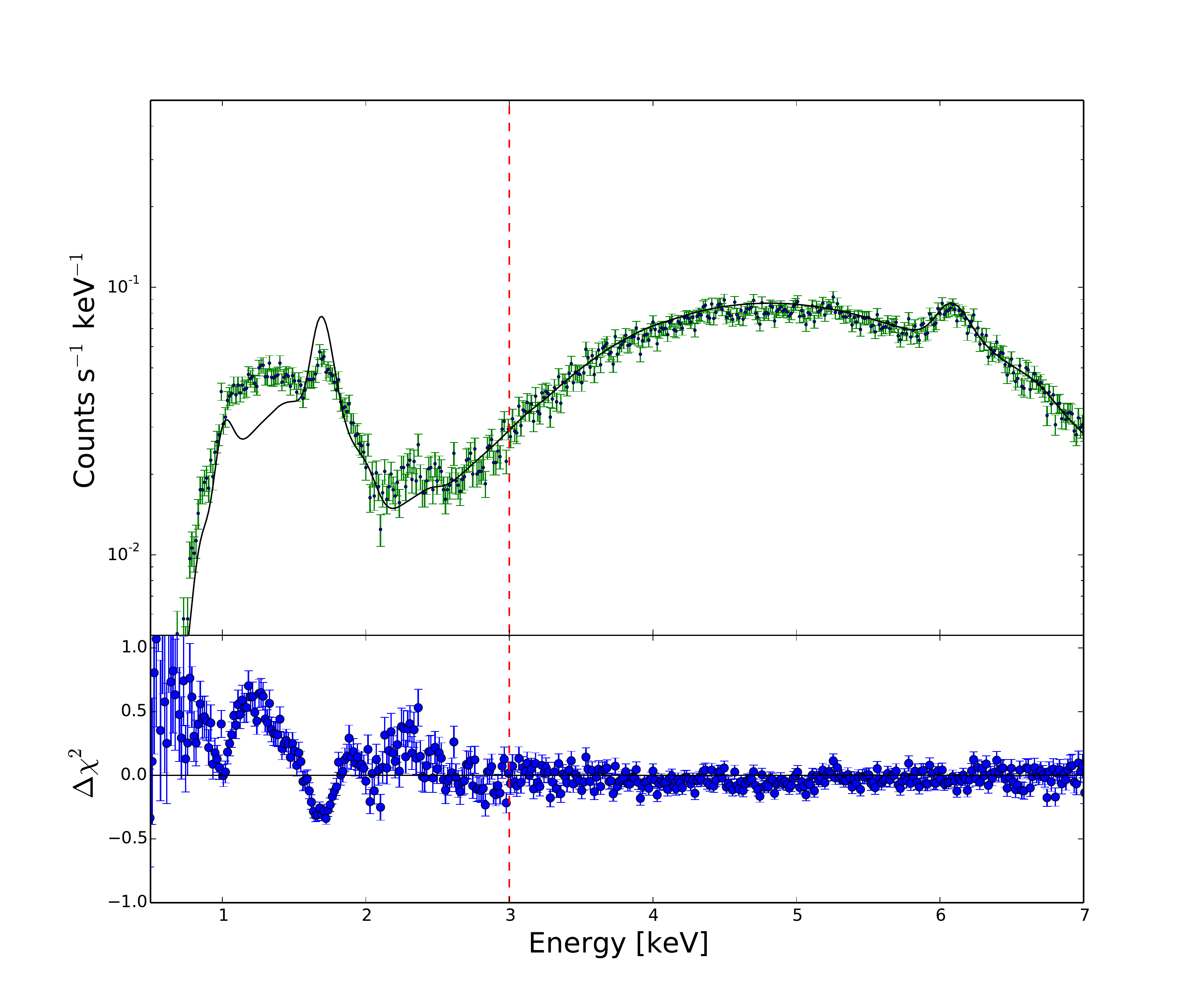}}
\caption{The spectral fits to the 2000 short and normal frame time, 2005 and 2015-2017 observations. Only energies above the red line at 3 keV are included in the fit}
\label{AGN:fig:AGNfits}

\end{figure*}

\begin{table*}
\centering
\caption{Results of our fits to the the 2000, 2005, and 2015-2017 {\it Chandra} ACIS spectra with the source model described in the text. Parameters directly obtained from the fit are shown with the 90\% confidence interval.}
\begin{tabular}{cccccccccc}
\hline \hline
Year & $\alpha$ \textsuperscript{a} & $f_{\rm pileup}$ \textsuperscript{b} & $N_{\rm H}$ \textsuperscript{c} & $\Gamma$ & $L_{2-10 \rm keV}$ & $\mu_{\rm iron-K\alpha}$ \textsuperscript{d} & EW$_{\rm iron-K\alpha}$ \textsuperscript{e} &  $\tau_{edge}$ \textsuperscript{f} & $\chi^2$ \\ 
& & & ($10^{22}$ cm$^{-2}$) & & ($10^{44}$ erg s$^{-1}$)  & (keV) & (eV) & & \\ \hline
 2000 & $0.55 \pm 0.05$ &  0.19 & $19.2 \pm 4.3$ & $1.42 \pm 0.12 $ & $2.0 \pm 0.2 $& $6.40 \pm 0.01 $ & 219 & $0.32 \pm 0.15$ & $330/320$\\
 2005 &  0.55 \textsuperscript{h} & 0.19 & $19.1 \pm 3.9$ & $1.30 \pm 0.04$ & $ 2.2 \pm 0.2 $ & $6.43 \pm 0.01 $ & 149 & $0.40 \pm 0.23 $ &  $335/321$ \\
 2015-2017 & 0.55 \textsuperscript{h} &  0.17 & $20.7 \pm 3.3$ & $1.23 \pm 0.04$ & $1.9 \pm 0.2 $ & $6.44 \pm 0.01 $ & 150 & $0.35 \pm 0.10$ & $412/383$ \\ \hline

\label{AGN:tab:hardemission} 

\end{tabular} \\
\raggedright{\setlength\parindent{0em}
\textsuperscript{a} The grade migration parameter. \\
\textsuperscript{b} The fraction of piled events. \\
\textsuperscript{c} The column density to the nucleus. \\
\textsuperscript{d} The median rest-frame energy of the iron-K$\alpha$ line. \\
\textsuperscript{e} The equivalent width of the iron-K$\alpha$ line. \\
\textsuperscript{f} The optical depth of the absorption edge at rest-frame energy 7.2 keV. \\
\textsuperscript{h} This parameter was fixed, see text for details. \\

}
\end{table*}

We have defined a circular extraction region with a radius of 3 arcsec, centered on the AGN. As a local background, we used an annulus around the source between 6 and 10 arcsec. The source and background regions, as well as the response files, were extracted from each ObsID with the CIAO tool \textit{specextract}. All observations in 2005 and between 2015 and 2017 were combined with \textit{combine\_spectra}. The 2005 and 2015-2017 observations were all taken with the ACIS-I detector, and thus we did not have to worry about different detector types. After extracting, we fit the spectra with the CIAO fitting and modeling tool \texttt{Sherpa} \citep{Freeman2001}. 

As previously mentioned, ObsID 1707 is unique because it is the only short frame time observation of the AGN, and therefore is not piled up. For ease of reference, we will refer to ObsID 1707 as the 2000 SF (short frame time) observation, and to ObsID 360 as the 2000 NF (normal frame time) observation. As our analysis in section \ref{AGN:subsec:PSF} has shown, there is moderate to strong pileup in this source which needs to be accounted for in the spectra of the normal frame time observations. To this end, we have used the pileup model \texttt{jdpileup} \citep{Davis2001}.

Several parameters are used in \texttt{jdpileup} to parametrize the amount of pileup in the spectrum. The most important of these are 1) $n$, describing the number of 3x3 islands in which pileup is applicable, 2) $f$, the fraction of the total flux of the spectrum that is inside the pileup region, and 3) $\alpha$, a parametrization of grade migration, where if $N$ photons are piled together, the chance that they are registered as a single photon with their combined energy is given as $\alpha^{N-1}$. 

The PSF simulations from the previous section have shown that pileup significantly affects the data up to about 1.3 arcsec radius.  We estimate the parameter $f$ from the 2.2-8.0 keV flux image of the 2000 SF observation. We calculate the flux ratio of the flux within the estimated pileup radius of 1.3 arcsec, and 3 arcsec, the total size of the extraction region. From the image we measure $f = 0.93$.   Although the PSF is a bit more extended in 2005 compared to 2000, we assume this will not significantly alter the value of $f$ and we have chosen to leave this parameter fixed.

The parameter $n$ represents the number of 3x3 pixel islands in the observation where pileup is significant. After the example in \cite{Davis2001}, we have counted the number of full pixels in a 1.3 arcsec circle and divided this number by 9. We count 14 pixels, and therefore we set $n=1.55$.

Next, we consider the spectral model. Previous modeling of the AGN by \cite{Young2002} and \cite{Reynolds2015} shows that the AGN spectrum is a heavily absorbed power law. Because of the heavy absorption, there are virtually no counts coming from the AGN below 2 keV. On top of the power law, there is a fluorescent iron-K$\alpha$ emission line in the spectrum, around 6.4 keV rest-frame energy. This line can be modeled either with a Gaussian, or with a more detailed Compton-thick reflection model. While the reflection model is the more physical of the two, it is difficult to fit this to the \textit{Chandra} data, where we are restricted to the 0.5-7.0 keV energy range, below most of the Compton reflection features. In fact, \cite{Reynolds2015} have attempted to fit the reflection model to the \textit{XMM-Newton} data and find that the reflection model does not significantly improve the fit over a Gaussian emission line. We therefore use a Gaussian to describe the fluorescent emission line. Beyond the iron-K$\alpha$ line, \cite{Young2002} note the presence of an absorption edge at a rest-frame energy of around 7.2 keV, or 6.8 keV observed energy. We therefore include this component in our model, and we have fixed the treshold energy of the absorption edge to 6.8 keV.

The presence of pileup in the spectrum means that we must consider the full \textit{Chandra} ACIS energy range, and not just photons above 2 keV, where the AGN is bright. This is because the soft X-ray photons will affect higher energies by becoming piled. This presents a problem, as the soft X-ray photons come from an extended region rather than the AGN. Therefore, the parameters of the pileup model are not necessarily valid. We have attempted to circumvent this problem by including the model components of the soft X-ray emission in the total model, but \textit{not} fitting the data below 3 keV. The soft X-ray emission is modeled as a second, unabsorbed power law as well as 3 fluorescent emission lines, modeled with Gaussians \citep[see][]{Young2002}. Given that this model is not fit to the data, we do not consider the fit values that we obtain for the soft X-ray model components as physically accurate. Rather, the soft X-ray model components exist purely to model how the soft X-ray band of the spectrum affects the shape of the spectrum in the hard X-ray band due to pileup. During the fitting process, we have tweaked the parameters of the soft X-ray model components, in particular the amplitude of the emission lines, such that they do not deviate too far from the soft X-ray data. 

Lastly, we take into account the contribution of the thermal ICM. We have modeled the ICM with a thermal \texttt{APEC} model, absorbed by the galactic absorption. We fit this model to the background annulus between 0.5 and 4 keV. At energies beyond 4 keV, the PSF of the AGN is so wide that the hard X-ray photons from the AGN significantly contribute to the background annulus. For each time period, we determine the abundance and temperature with a fit to the background annulus. We then freeze the temperature and abundance in the fit to the source spectrum, allowing only the normalization to vary. The temperature and abundance are slightly different between 2000 and the later time periods. This can be explained by the fact that the 2000 observations were taken with ACIS-S and the later observations with ACIS-I. The changes are of order 5-10\%, and the ICM temperature and abundance for all spectra are around $kT = 3.5$ keV and $Z = 0.55$ Z $_\odot$. 

The final model is then as follows: \texttt{ABS1 * (ICM + POW1 + 3*emLINES + ABS2 * EDGE * (POW2 + emLINE))}. This model is used as input to the \texttt{jdpileup} model for all spectra except the 2000 SF observation. The galactic absorption is set to $3.1 \times 10^{21}$ cm$^{-2}$ \citep{Dickey1990}. The FWHM of each emission line is set to 0.07 keV, roughly the expected FWHM of \textit{Chandra}. We took the energies of the soft X-ray emission lines from \cite{Young2002}, and left the energy of the iron-K$\alpha$ emission line as a free parameter. We have also linked the photon index of POW1 to the photon index of POW2, which is what is expected if the soft X-ray emission is scattered AGN light.

We have fit the 2000 SF and 2000 NF spectra simultaneously. Because these were taken days apart, we have assumed that parameters such as the photon index and the absorbing column have stayed the same in this time period. Simultaneously fitting the same model to these two spectra helps to further constrain the pileup model. The 2005 and 2015-2017 spectra were fit separately, giving us 4 spectra and 3 time periods in total. 

Our best fits to the spectra are shown in Fig. \ref{AGN:fig:AGNfits}, and the results of those fits are given in Table \ref{AGN:tab:hardemission}. During the fitting process we noticed that the grade migration parameter $\alpha$, when left as a free parameter, takes on a value of 1 when fitting the 2005 and 2015-2017 spectra. Because more grade migration hardens the spectrum, it is likely that there is some degeneracy between $\alpha$ and the photon index $\Gamma$, which makes it difficult to find the appropriate value of $\alpha$. In the 2000 spectra this degeneracy is easier to break, because the SF and NF spectra are fit simultaneously. We therefore fixed the grade migration parameter in these observations to 0.55, the value found from the fit to the 2000 observations. 

Furthermore, there seems to be no significant change in the spectra between the three different time periods. The photon index is lower in the 2005 and 2015-2017 observations, but given the degeneracy between $\alpha$ and $\Gamma$, it is unclear whether this is a real trend. The photon indices we find are also broadly consistent with the photon index $\Gamma = 1.43 \pm 0.11$ that \cite{Reynolds2015} find for the XMM-Newton data, using the \texttt{ICM + cABS(PL + emLINE)} model. The photon index is found to be higher, around $1.60$, in the \textit{NuSTAR} data, but this is somewhat model dependent. In particular, adding reflection and wind model components changes the way the region around the iron-K$\alpha$ line and beyond is modeled, and this can shift the photon index of the entire spectrum. 

One particular challenge of pileup models is that it can be difficult to distinguish between a source with a high flux/high pile-up fraction, and low flux/low pile-up fraction. For each of the fits, we have calculated the pileup fraction in \texttt{Sherpa} with the command \textit{print(get\_pileup\_model())}. The pileup fraction for all observations is between 0.17 and 0.19, comparable to the pileup fraction of 0.22 found in Section \ref{AGN:subsec:PSF}.  Additionally, the luminosity in 2015-2017 is about 95\% of the 2000 luminosity, consistent with the 90\% estimate we used to simulate the data with \texttt{marx}. 

We have used the \texttt{Sherpa} tool \textit{sample\_flux} to estimate the 90\% confidence interval on the AGN luminosity. We  find that for all time periods, the error is on the order 10\%. 

The \textit{XMM-Newton} data has significantly higher absorbing column ($3.0 \times 10^{23}$ cm$^{-2}$ ) than the \textit{Chandra} observations from 2005. Because these observations were taken close together in time, we would expect the absorbing column to be roughly the same value. We tried to re-fit the 2005 \textit{Chandra} spectrum by fixing the absorbing column to the \textit{XMM-Newton} value, but were unable to find a good fit with this constraint. Most of Chandra exposures in 2005 were taken in February, with only ObsID 6252 being taken later, in September. This is also the ObsID that is closest in time to the \textit{XMM-Newton} observation date in October. To see whether the absorbing column changed between February and September, we fit the spectral model to ObsID 6252 separately to look for signs of a higher absorbing column. However, we find a value that is consistent with that of the total 2005 spectrum, $N_H = (17.6 \pm 2.2) \times 10^{22}$ cm$^{-2}$. We therefore find no indication in the \textit{Chandra} data that the absorbing column increased towards the \textit{XMM-Newton} observation date.

The intrinsic luminosity that we find from simultaneously fitting the 2000 SF and 2000 NF spectra is $ L_{2-10 \rm keV} = 2.0 \times 10^{44}$ erg s$^{-1}$. This is consistent with the value of $L_{2-10 \rm keV}  = 1.9 \times 10^{44}$ erg s$^{-1}$ that \cite{Young2002} find, using the 2000 SF observation. Furthermore, the 2005 \textit{Chandra} luminosity is consistent with the \textit{XMM-Newton} luminosity.

\subsection{Swift XRT observations}

\label{AGN:subsec:swift}

\begin{table}
\caption{The {\it Swift} XRT observations of the field of Cygnus A, in Photon Counting Mode. ObsIDs between dashed lines were combined.}
\begin{tabular}{cccc}
\hline \hline
Date                & Observation ID & Exposure & Distance from centre \\
(UTC)               &                & (ks)      & (arcsec) \\
\hline
2006-01-23T05:08 & 00035024002    & 5.1     & 103      \\
\hdashline
2007-03-06T04:26 & 00036397001    & 9.0     & 98       \\
2007-03-08T00:57 & 00036397002    & 12.4    & 84       \\
\hdashline
2008-08-23T15:08 & 00036397003    & 5.6     & 118      \\
2008-08-27T04:19 & 00036397004    & 3.7     & 139      \\
\hdashline
2013-02-17T20:08 & 00080235002    & 2.0     & 118      \\
2013-02-27T23:58 & 00082067001    & 1.9     & 399      \\
2013-03-01T16:01 & 00080235003    & 1.9     & 95       \\
\hline \\

\end{tabular}

\label{AGN:tab:SwiftObs}
\end{table}

To be able to study the long-term behaviour of the nuclear region, we have searched for other X-ray observations of Cygnus A between 2000 and 2015. {\it Swift} XRT has observed Cygnus A eight times between January 2006 and March 2013. Two observations, on the 17th of February and 1st of March 2013, correspond exactly to \textit{NuSTAR} observation dates. The full observation log is given in Table \ref{AGN:tab:SwiftObs}. 

For each of these observations, obtained in Photon Counting Mode, we have created a spectrum and associated response files. This was done using \texttt{HEASoft} version 6.24, and the associated CALDB for \textit{Swift}.  Because the FWHM of the XRT PSF is 18 arcsec, the AGN is not clearly resolved inside the larger Cygnus A X-ray environment. Therefore, any spectrum extracted from the Cygnus A region is significantly contaminated by emission from the surrounding ICM, which is relatively hot and bright. This is in principle no different from \textit{NuSTAR}, but there are additional problems for \textit{Swift} XRT: the count rate of the observations, around 0.56 counts/sec, is above the approximate pileup treshold count rate of 0.5 counts s$^{-1}$ \footnote{http://www.swift.ac.uk/analysis/xrt/pileup.php}. The data is therefore likely piled up. Additionally, \textit{Swift} is more sensitive to soft X-ray photons than \textit{NuSTAR}, which means the contribution of the ICM to the spectrum is stronger. 

While the \textit{Swift} software has tools to correct for pileup by excising the center pixels of the observation, these tools assume that the extracted region is a point source on a constant background. Because of the ICM that the AGN is embedded in, this is not the case here and therefore any pile-up correction will be inaccurate. 

To minimize ICM contamination, we chose a relatively small extraction circle of 18 arcsec, centered on the brightest pixel of each ObsID. We combined the spectra from 2 ObsIDs in March 2007, 2 in August 2008, and 3 in February/March 2013. Together with the 2006 data point, this makes 4 data points in total. 

All of the \textit{Swift} spectra have a strong peak in the soft X-ray band, between 0.5 and 4 keV, and a peak in the hard X-ray band, between 4 and 10 keV. We interpreted the soft X-ray peak to be the ICM and have therefore excluded energies below 4 keV, where the ICM is the brightest. We fit an observed power law to the spectrum between 4-10 keV and use that to determine the observed luminosity. We note that, given the presence of pile-up, and the potentially still significant presence of contaminating ICM, we do not regard the obtained fit values for the photon index and the absorbing column as physically accurate. Instead, the model serves as a way to determine the observed, rather than the intrinsic, luminosity from the spectra. 

After obtaining the observed luminosities, we estimated the intrinsic luminosities by applying a few different corrections. Firstly, we applied a pileup correction. Because the count rate is slightly above the pileup count rate treshold, we assume a small pileup fraction of 10\%. This rough estimate is based on the fact that the count rate in our Swift spectra, 0.56 counts s$^{-1}$, is roughly 10\% higher than the pileup treshold count rate, 0.5 counts s$^{-1}$. Secondly, we estimated the ICM contribution by fitting an absorbed thermal model to the spectra between 0.5-2.0 keV, and calculating the luminosity of this thermal model between 4-10 keV. The estimated 4-10 keV ICM luminosity within the extraction region is $\sim 2.5 \times 10^{43}$ erg s$^{-1}$. We subtracted this from the pileup-corrected observed luminosity. Thirdly, we increased the luminosity by 5\% to account for the difference between the 2-10 keV and 4-10 keV band. Finally, we de-absorbed the luminosities, by assuming a photon index $\Gamma=1.4$. In the 2006 and 2013 observations, we assumed the absorbing column to be the same as in the 2005 {\it Chandra} and 2013 {\it NuSTAR} observations respectively. In the 2007 and 2008 observations, we assumed a range in absorbing column between $1 - 3 \times 10^{23}$ cm$^{-2}$, which adds an uncertainty of 40\% to the calculated intrinsic luminosity.

\subsection{X-ray observations before 2000}
\label{AGN:subsec:hist}

To measure the long-term X-ray behaviour of Cygnus A, we searched the literature for X-ray observations before 2000. Cygnus A has been observed by several X-ray telescopes before \textit{Chandra}, although in these observations, the AGN was not spatially resolved from the ICM. Given the lower spatial and spectral resolution of these instruments, and the relative brightness of the ICM, the measured intrinsic AGN luminosities are a lot more uncertain.

Cygnus A was observed with the European X-ray Observatory Satellite (\textit{EXOSAT}) in 1985 \citep{Arnaud1987}. A fit to the data with a thermal component and an absorbed power law yielded an intrinsic 2-10 keV luminosity of $2.8 \times 10^{44}$ erg s$^{-1}$, while the thermal component has a  luminosity of $3.4 \times 10^{44}$ erg s$^{-1}$.

In 1991, Cygnus A was observed with the \textit{Ginga} satellite \citep{Ueno1994}. Like with the \textit{EXOSAT} observations, the authors find that the best description of the data is provided by a model consisting of 1) a thermal and 2) an absorbed power law component. The intrinsic luminosity was found to be much higher than during the EXOSAT observation, about $5.5 \times 10^{44}$ erg s$^{-1}$, and a thermal luminosity of $5.8 \times 10^{44}$ erg s$^{-1}$. However, the Large Area Counter instrument on board {\it Ginga} that was used to observe Cygnus A has a PSF with a FWHM of $1\degree .1 \times 2\degree.0$ \citep{Turner1989}. With such a large field of view, it measures the spectrum of the entire cluster, and can not be used to make a reliable measurement of the AGN luminosity. We therefore do not include the {\it Ginga} luminosity in our light curve.

Finally, Cygnus A was observed by the Advanced Satellite for Cosmology and Astrophysics (\textit{ASCA}) in May and October 1993. The data were analysed by \cite{Sambruna1999}. Fitting a thermal model plus absorbed power law to these data, they found that the nuclear flux increased by a factor of more than 3 in 5 months, while the absorbing column increased from  $1.1 \times 10^{23}$ cm$^{-2}$ to  $1.5 \times 10^{23}$ cm$^{-2}$. Because we were unable to reproduce the conversion from observed to intrinsic luminosity listed in that paper, we retrieved the archival data from the HEASARC data archive and re-analysed the spectra. We selected the pre-extracted spectrum of the entire field that is included in the standard data products, and used the spectra of the Solid State Imager (SIS) instrument in Bright mode.

Comparing {\it ASCA} observations, we find that the 2-10 keV count rate is about 10\% higher in October than in May. We freeze the photon index to 1.8, the value found by \cite{Sambruna1999}, and leave the other parameters free. We fit the spectra between 3 and 9 keV and find intrinsic luminosities for the non-thermal component of $3.1 \times 10^{44}$ erg s$^{-1}$ and $3.0 \times 10^{44}$ erg s$^{-1}$ for May and October, respectively. Interestingly, the absorbing column is also much higher than what was found by \cite{Sambruna1999}, $2.6 \times 10^{23}$ cm$^{-2}$ and $3.1 \times 10^{23}$ cm$^{-2}$ respectively.  This is likely because we used 3 keV instead of 4 keV as a lower limit, which allows for a better restriction of the absorbing column.

\subsection{The AGN light curve}

\begin{figure*}

   \includegraphics[width=0.9\textwidth]{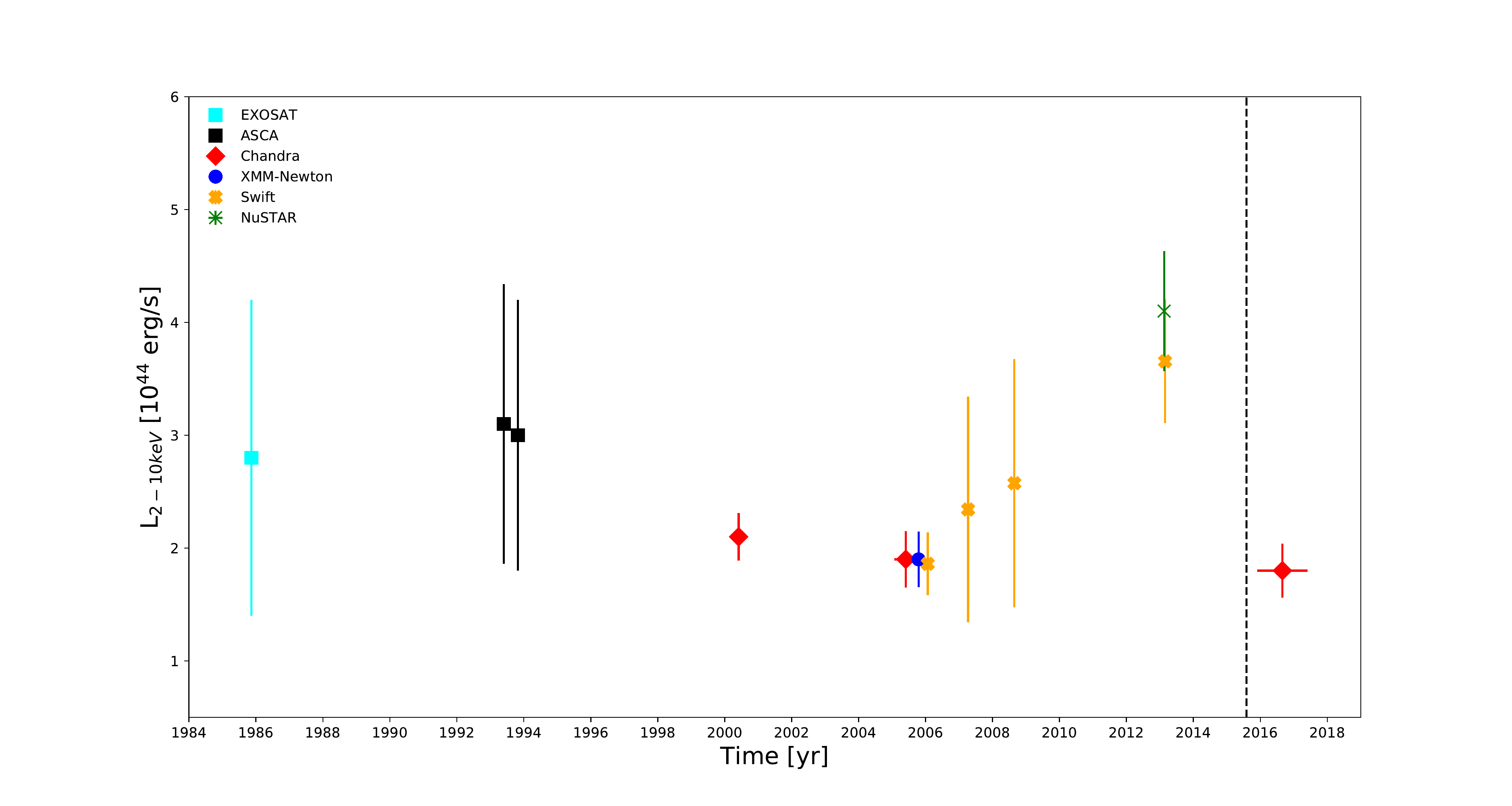}

	\caption{The intrinsic 2-10 keV luminosity of the AGN of Cygnus A over time, as measured by \textit{EXOSAT}, \textit{ASCA}, \textit{Chandra}, \textit{XMM-Newton},  \textit{NuSTAR} and \textit{Swift}. The dashed line indicates the date of first discovery of Cygnus A-2 with the VLA.}
    
       \label{AGN:fig:timeflux}

\end{figure*}

We have combined all the available luminosity data on the AGN of Cygnus A into a single light curve in Fig. \ref{AGN:fig:timeflux}. Shown are the {\it Chandra} data from section   \ref{AGN:subsec:spectral}, the \textit{Swift} data from section \ref{AGN:subsec:swift}, the 
{\it EXOSAT} and {\it ASCA} data from section \ref{AGN:subsec:hist}, and the {\it XMM-Newton} and {\it NuSTAR} data from \cite{Reynolds2015}.

We note that the errors on the luminosities are, in some cases, rather uncertain. No errors on the luminosity are reported for the {\it EXOSAT} data point in \cite{Arnaud1987}, nor for the {\it XMM-Newton} and {\it NuSTAR} data points in \cite{Reynolds2015}. For {\it EXOSAT}, we based our uncertainty estimate on the uncertainty in the absorbing column, as well as adding 25\% systematic uncertainty because of the fact that the AGN is not resolved within the larger environment, and the limitations of the simple two-component model that is fit to this data. For the {\it ASCA} data, which we have re-analysed in this work, we used the {\it sample\_flux} tool in \texttt{Sherpa} to obtain 90\% confidence intervals. For the {\it Chandra} data, we similarly used {\it sample\_flux} to estimate the 90\% confidence interval. For the {\it XMM-Newton} and {\it NuSTAR} data, we based our estimate on uncertainty in the the absorbing column as well as a 5\% systematic error. For {\it Swift} XRT, we based our estimate on the uncertainty in the absorbing column and a 25\% systematic error. The systematic error in the case of {\it Swift} is large because of the assumptions we had to make when converting observed luminosity to intrinsic luminosity.

Although there are some hints that the average luminosity was higher before 2000, this increase is not significant because of the large error bars on the {\it EXOSAT} and {\it ASCA} luminosities. All data points are consistent with a luminosity of $2 \times 10^{44}$ erg s$^{-1}$, except for in 2013, when a statistically significant higher luminosity was observed by both {\it NuSTAR} and {\it Swift}. We discuss potential reasons for this increase in sections \ref{AGN:subsec:xraylc} and \ref{AGN:subsec:TDE}.

\subsection{Wind signatures in the Chandra data}

\label{AGN:subsec:wind}

\begin{figure}
   \includegraphics[width=0.5\textwidth]{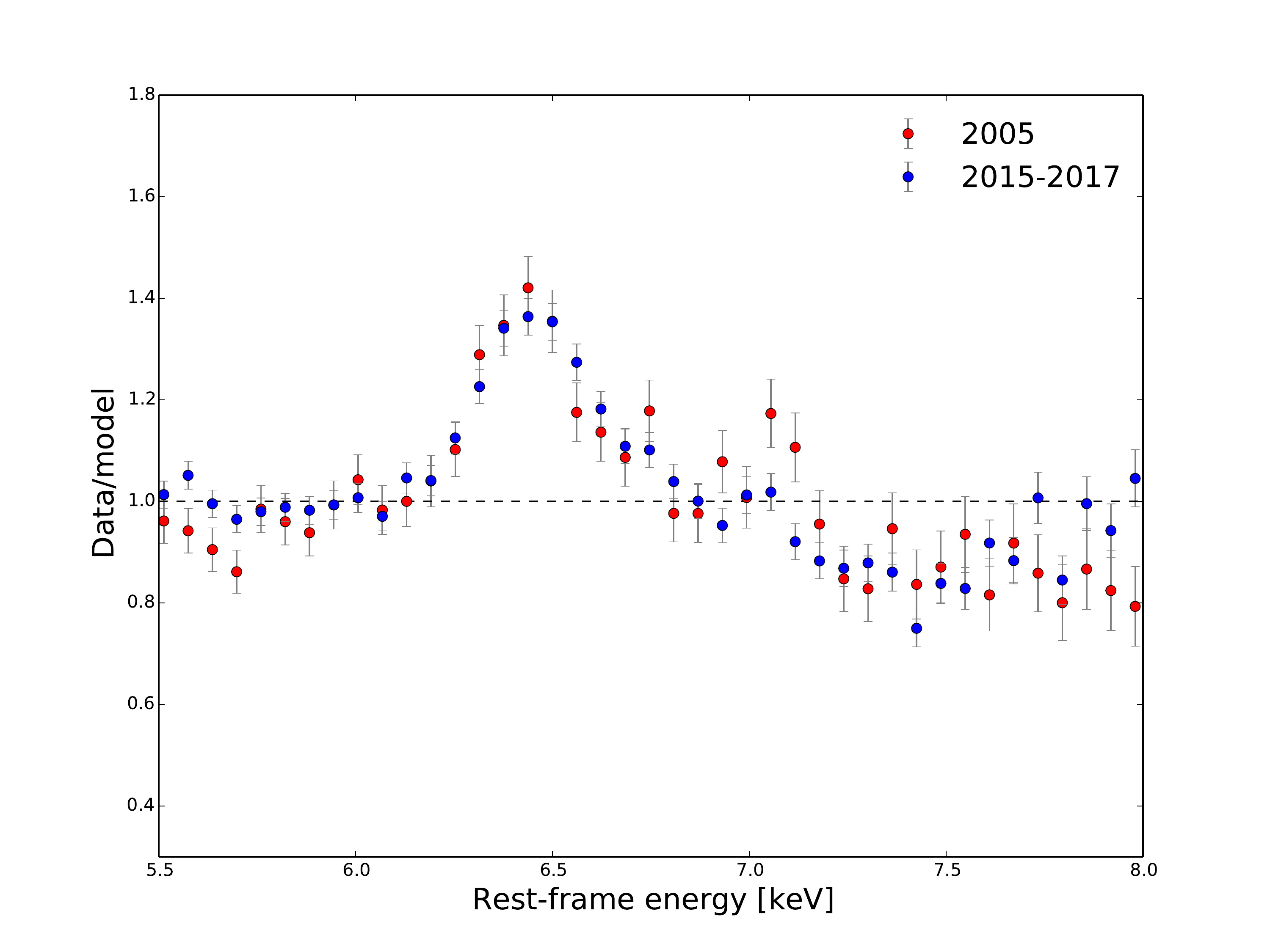}

  \caption{The \textit{Chandra} ACIS  2005 and 2015-2017 ratio spectra. The horizontal axis shows the Cygnus A rest-frame energy. The data have been divided by a piled, absorbed powerlaw, leaving the iron K line and the absorption edge at 7.2 keV.}
  
    \label{AGN:fig:ironK}

\end{figure}

\cite{Reynolds2015} previously detected a fast and ionized wind in the 2013 {\it NuSTAR} spectrum. The presence of this wind was deduced from a P Cygni-like spectroscopic feature around the iron-K$\alpha$ line: a redshifted emission component that moves the centroid of the iron-K$\alpha$ line to lower energies, and a blueshifted absorption component of the power law continuum. The 2005 {\it XMM-Newton} spectrum showed no evidence of such an outflow, leading to the conclusion that the wind appeared somewhere between 2005 and 2013.

To compare the the spectral features around the iron-K$\alpha$ line between the 2005 and 2015-2017 \textit{Chandra} observations, we have made ratio plots by dividing the data by the piled up power law model. These are shown in Fig. \ref{AGN:fig:ironK}.  Since the fluorescent iron K$\alpha$ line and the absorption edge are not included in the model used for the divisor, their structure can be seen in the ratio spectra. A visual comparison between the two spectra shows that the centre of the iron K line has not redshifted. In fact, as Table \ref{AGN:tab:hardemission} shows, it appears that the line centroid has slightly {\it blueshifted} between 2000, 2005, and 2015-2017. During the {\it XMM-Newton} observation in 2005, the line centroid was at $6.39 \pm 0.01$ keV, lower than the 2005 {\it Chandra} value of $6.43 \pm 0.01$ keV .

We think that the most likely explanation for the shift is uncertainties in the gain calibration, and the different chips that were used during each epoch. In the 2000 observations, the AGN was on the S3 chip. In the 2005 observations, the AGN was on I1, with the exception of ObsID 6252, where it was on I3. In all the 2015-2017 observations, the AGN was on I3. According to the {\it Chandra} Calibration Status Report from July 2017, the average uncertainty in detector gain is $0.3\%$. The uncertainty in gain calibration is lowest for the back-illuminated chips, but it can be up to $0.6\%$ on front-illuminated chips at higher focal plane temperatures and at high values of CHIPY. The 2005 observations were all taken at the standard focal plane temperature of $-119.7 \degree$, but some of the 2015-2017 observations had higher temperatures, of up to $-115 \degree$. Additionally, the nucleus is positioned at high CHIPY values on most observations. We believe that these factors, combined with the $0.01$ keV uncertainty from the fit, can account for the $0.4-0.6\%$ shift between 2000 and the later observations. 

However, in all {\it Chandra} observations the line centroid is significantly higher than during the \textit{NuSTAR} observation, $\mu_{NuSTAR}= 6.34_{-0.04}^{+0.03}$. Therefore, based on the centroid of the iron-K$\alpha$ line, there is no evidence for a redshifted wind component in the {\it Chandra} spectra.

The absorption edge can clearly be seen in both the 2005 and 2015-2017 spectra. \cite{Reynolds2015} speculate that the absorption edge could be the result of iron-K$\alpha$ absorption in an ionized wind. The fact that the ionized wind is not seen in the \textit{XMM-Newton} data could then be explained by the fact that the absorbing column was significantly higher during that observation, in October 2005. As discussed in section \ref{AGN:subsec:spectral}, the \textit{Chandra} observations from 2005 show no evidence of a higher absorbing column. Fitting the spectral model to individual ObsID 6252, from September 2005 does not return a higher absorbing column than the average 2005 value. Therefore, if the absorbing column was higher during the \textit{XMM-Newton} observation, it must have increased from roughly $2-3 \times 10^{23}$ cm$^{-2}$ in the month between the observations.  This would be a relatively rapid change, especially considering the absorbing column is around the same value in all \textit{Chandra} and \textit{NuSTAR} observations.

Furthermore, supposing that the fast wind was present in all the \textit{Chandra} data, and is simply related to the amount of absorbing material in our line of sight, it is puzzling why the centre of the iron-K$\alpha$ emission line seen with \textit{Chandra} is inconsistent with that of \textit{NuSTAR}. This alone seems to be the strongest evidence that the fast wind is not present in either 2005 or in 2015-2017. However, this still leaves the exact nature of the absorption edge as an open question.

\section{Discussion}
\label{AGN:sec:disc}

\subsection{AGN variability in Cygnus A}
\label{AGN:subsec:xraylc}

AGN are known to be variable on a multitude of timescales, ranging from hours to years \citep[e.g.][]{Mushotzky1993, Markowitz2003, Gonzalez2012}. In the light curve in Fig. \ref{AGN:fig:timeflux}, we observe that the X-ray luminosity doubled between 2005 and 2013, and then halved again between 2013 and 2015. In this section, we discuss whether the rise and fall in luminosity can be explained by intrinsic variability of the AGN.

Although most observations have roughly the same duration, on the order of kiloseconds, for the \textit{Chandra} 2015-2017 observations we combined observations over a significantly longer time span. To verify that we are not missing significant variability within that time period, we looked at the 2-10 keV count rates of all the ObsIDs that make up the combined 2015-2017 observation. The count rate is steady with variations of order 10\% around the mean. From these observations, there is therefore no evidence that the AGN is highly variable on the timescale of weeks and months. To explain the increased luminosity seen by \textit{NuSTAR} we would therefore need to turn to long-term variability.

An extensive study of long-term AGN variability in the 2-10 keV band was carried out by \cite{Sobolewska2009}. In the 10 AGN that were studied, a change in luminosity by a factor of 2 seems to certainly be within the realms of possibility. However, this also depends strongly on the properties of the AGN itself. Stochastic modeling of AGN variability shows that the characteristic decay timescale is larger if the black hole mass is larger \citep{Kelly2011}. For the AGN of Cygnus A, with a black hole mass $M_{BH} = 2.5 \times 10^{9}$ M$_\odot$ \citep{Tadhunter2003}, this timescale is on the order of $\sim 50-1000$ days. This means that it could take years for the AGN luminosity to decay back to its mean value. The time between the \textit{NuSTAR} and earliest \textit{Chandra} observations is more than 2.5 years, so in principle this could have been enough time for the AGN to settle back to its mean luminosity. 

\cite{Soldi2014} carried out a long-term AGN variability study with the \textit{Swift} Burst Alert Telescope (BAT). For this study, 66 month lightcurves (December 2004- May 2010) in the 15-100 keV band were used. The sample includes Cygnus A. The variability is parametrized as  $S_{V}$, the sample standard deviation expressed as a percentage of the mean. For Cygnus A, $S_{V} = 31_{-5}^{+4} $\%, compared to the average $\langle S_{V} \rangle = 24 \pm 1$ for the full sample of 11 radio-loud galaxies. 

However, how $S_{V, 14-100 keV}$ translates to variability between 2-10 keV is not directly obvious. At energies above 10 keV, reflection contributes to the total emission, as evidenced by the Compton hump in the \textit{NuSTAR} spectrum around 20-30 keV \citep{Reynolds2015}. The reflection component is expected to be constant because any variability will be smeared out by the light travel time towards the reflector \citep{Sobolewska2009, Bianchi2009}. Because the reflection component decreases in strength towards lower energies, the variability at 15-100 keV should be lower than at 2-10 keV. 

The observations of Cygnus A before 2000, with \textit{EXOSAT} and \textit{ASCA} show luminosities that are consistent with what we have measured with more recent missions. However, as mentioned before, the relatively low spatial and spectral resolutions make it more difficult to accurately separate the thermal from the non-thermal emission. Fitting a single thermal model to the entire cluster emission would likely leave a hard residual in the spectrum of gas that is hotter than the mean. This hot emission might have been attributed to the AGN, leading to higher measured luminosities. Additionally, the absorbing column is not always well-constrained. This means the luminosities measured with these missions are uncertain and that they are consistent with both the idea of a relatively steady AGN, or an AGN doubling or perhaps even tripling its luminosity within a few years.

Given the stochastic nature of AGN variability and the long-term studies of AGN variability at X-ray wavelengths mentioned above, we can not rule out that the 2013 peak in the Cygnus A light curve is simply the nucleus going through a temporary high-accretion phase. 

\subsection{The X-ray luminosity upper limit of Cygnus A-2}
\label{AGN:subsec:orcyga2}

In section \ref{AGN:subsec:cygA2} we have estimated an upper limit to the X-ray luminosity of Cygnus A-2, by looking for coincident excess emission in the 2015-2017 {\it Chandra} observations. The observed 2-10 keV luminosity of the primary AGN in the 2015-2017 {\it Chandra} observation is $\sim 8 \times 10^{43}$ erg s$^{-1}$, which gives an upper limit to the observed luminosity of Cygnus A-2 of  $\sim 4 \times 10^{42}$ erg s$^{-1}$. How this translates to intrinsic luminosity depends on the absorbing column and spectral shape of Cygnus A-2. If we assume the spectrum to be an absorbed power law and take the absorbing column to Cygnus A as an upper limit, our upper limit estimate of the 2-10 keV intrinsic luminosity is $\sim 1 \times 10^{43}$ erg s$^{-1}$. 

One of the scenarios that \cite{Perley2017} proposed is that Cygnus A-2 is a steadily accreting secondary AGN. If the radio emission comes from accretion mechanisms, it implies a highly variable AGN, because of the non-detection of Cygnus A-2 in earlier radio observations. Based on the luminosity of the infrared point source that coincides with Cygnus A-2 \citep{Canalizo2003}, this black hole would be accreting at far below the Eddington limit. This means it is expected to lie on the Fundamental Plane for sub-Eddington accreting black holes, given in \cite{Plotkin2012} as 

\begin{equation}
\label{AGN:eq:fp}
\log{L_{\rm X}} = (1.45 \pm 0.04) \log{L_{\rm R}} - (0.88 \pm 0.06) \log{M_{\rm BH}} - 6.07 \pm 1.10 .
\end{equation}

In Eq. \ref{AGN:eq:fp}, $L_{\rm X}$ is the X-ray luminosity between 0.5 and 10 keV,  $L_{\rm R} = (\nu L_{\nu})_{\rm 5 GHz}$ the radio luminosity, and $M_{\rm BH}$ the mass of the black hole in solar mass. If we assume a spectrum with a photon index of $\Gamma = 1.4$, similar to the primary AGN, $L_{0.5 - 10}  \sim 1.2 \times 10^{43} $ erg s$^{-1}$. 

Although no 5GHz observations of Cygnus A-2 were reported on by \cite{Perley2017} we assume $F_{\rm 5 GHz} = 3.5$ mJy by extrapolating the optically thin spectral model in their Fig. 2. This results in a radio luminosity $L_{R} = 1.3 \times 10^{39} $ erg s$^{-1}$, and a lower limit to the black hole mass of $4 \times 10^{8}$ M$_\odot$. 

\cite{Perley2017} have previously noted two constraints to the upper limit of the Cygnus A-2 black hole mass: firstly, no dynamical disturbances are observed from IR spectroscopy at the location of Cygnus A-2, implying a mass significantly lower than the mass of the primary black hole \citep[$ \sim 2.5 \times 10^{9}$ M$_\odot$,][]{Tadhunter2003}. 

The intrinsic scatter of the FP is given in \cite{Plotkin2012} as $\sigma_{\rm int} = 0.07 \pm 0.05$ dex. This means the lower limit that we find can potentially be a little bit lower, but not by a significant amount, and is likely to be at least $10^8 M_\odot$. This is a significant fraction of the black hole mass of the primary AGN. Additionally, the $L_{\rm R} - M_{\rm BH}$ relationship \cite{Franceschini1998}, and the pre-flare upper limits to the radio flux density imply a black hole mass $M_{\rm BH} \sim 10^8 M_\odot$, which is inconsistent with our {\it lower} limit of $4 \times 10^{8}$ M$_\odot$.

Lastly, an estimate for the black hole mass can be obtained from the infared luminosity of Cygnus A-2 during the 2003 {\it Keck} observation \citep{Canalizo2003}. This is done by using the scaling relationship between the black hole mass, the 1\textmu m luminosity, and the width of the Paschen-$\beta$ broad emission line \citep[][equation 2]{Landt2013}. Although neither of these values is reported in the paper, the $\nu L\nu$ luminosity at 2\textmu m is given as $1.7 \times 10^{41}$ erg s$^{-1}$. Using this as an upper limit, and assuming a broad Pa$\beta$ line with a FWHM of 7500 km s$^{-1}$, implies a black hole mass of $3.2 \times 10^{7}$ M$_\odot$, an order of magnitude lower than we get from the X-ray estimate. However, it is unclear how long-term AGN variability plays into this scenario. It is possible that in 2003, Cygnus A-2 was still in the process of becoming 'active' and thus had not yet reached its full potential infrared luminosity.

Although none of the above arguments give us hard limits, the result is that the range of possible black hole masses for Cygnus A-2 is very narrow: it would require the X-ray luminosity to be at the upper limit, or for Cygnus A-2 to have an absorbing column significantly higher than $2 \times 10^{23}$ cm$^{-2}$, for the black hole mass to be at the lower end of the fundamental plane due to intrinsic scatter, and for the $L_{\rm Radio} - M_{\rm BH}$ correlation to be off by a factor of a few. Therefore, although \cite{Perley2017} favoured the explanation of Cygnus A-2 as an accreting black hole, the lack of X-ray flux seems to argue against this idea. We therefore turn to the alternative scenario of Cygnus A-2 as a TDE, which we discuss in the following section.

\subsection{Cygnus A-2 as a TDE}
\label{AGN:subsec:TDE}

As an alternative to a steady accretion scenario, Cygnus A-2 could have suddenly increased in brightness through a TDE. This would more naturally explain the absence of detectable X-ray emission in the 2015-2017 \textit{Chandra} observations: not all TDE's emit X-ray radiation, and when they do the X-ray emission typically fades on a scale of weeks to months \citep[e.g.][]{Auchettl2017}. While the TDE rate in an average galaxy is estimated to be on the order of $10^{-4}$ - $10^{-5}$ yrs$^{-1}$, analytical modeling shows that supermassive black hole binary systems may perturb stellar orbits around them enough to yield a significantly higher disruption rate, of up to $10^{-1}$ yrs$^{-1}$ \citep{Liu2013}. The recent discovery of a TDE in an ultra-luminous infrared galaxy by \cite{Tadhunter2017} gives some observational evidence that this may be the case. However, it is unclear whether the separation between the primary AGN and Cygnus A-2, a projected offset of 460 pc, is close enough to result in a significant perturbation of the stellar orbits.

Furthermore, we suggest that the peak in 2-10 keV X-ray luminosity in 2013 might be emission from that TDE. The $\sim 2.5$ years between {\it NuSTAR} and {\it Chandra} observations would have been more than enough for the X-ray emission to fade. Radio emission in TDE's has been observed for $\sim 2-3$ years after the disruption \citep[e.g][]{Bright2018b}, so an initial disruption in late 2012 or 2013 would be consistent with both X-ray and radio observations.

The difference in 2-10 keV luminosity in the Cygnus A light curve between 2013 and earlier observations is around $\sim 1-2 \times 10^{44}$ erg s$^{-1}$, but these intrinsic luminosities were calculated by using the absorbing column of the primary AGN, $N_{\rm H} \approx 20 \times 10^{22}$ cm$^{-2}$. If Cygnus A-2 were instead responsible for the increase in X-ray luminosity, the intrinsic luminosity would likely be lower, because the absorbing column would be a lot lower. The sample of TDE's studied in \cite{Auchettl2017} all have absorbing columns which are, although enhanced compared to their Galactic columns, not larger than $\sim 10^{22}$ cm$^{-2}$. Based on a range of reasonable absorbing column values, we therefore estimate the intrinsic luminosity to be $\sim (0.5 - 1) \times 10^{44}$ erg s$^{-1}$.

Most X-ray emitting TDE's radiate the majority of their energy away through soft, thermal X-ray emission from an optically thick disk, below 2 keV. The exception is the rare class of jetted TDE's, such as Sw J1644+57 \citep{Burrows2011, Zauderer2013} and Sw J2058+05 \citep{Pasham2015}. In jetted TDE's, the non-thermal X-ray emission in the 2-10 keV band can reach peak luminosities of $10^{47}$ erg s$^{-1}$.  Assuming the standard TDE light curve evolution of $t^{-5/3}$, 2-10 keV luminosities of order $10^{44}$ erg s$^{-1}$ could be reached after hundreds of days. If the TDE had ever reached a peak luminosity of $10^{47}$ erg s$^{-1}$, it would have been brighter than the rest of the galaxy by orders of magnitude, and it seems unlikely that it would not have been noticed by {\it Swift} BAT, which continuously monitors the full sky. However, the high luminosities in jetted TDE events are a result of the jet's direct orientation toward our line of sight. A slightly different viewing angle could potentially have lead to a lower X-ray peak luminosity more comparable to the X-ray luminosity of the primary AGN.

Alternatively, it has previously been shown in the literature that some thermal TDE's also have a non-thermal emission component in the 2-10 keV band, as in the TDE XMMSL1 J0740-85 \citep{Kawamuro2016,Saxton2017}. This power-law component is thought to be from the hot corona which Comptonizes the thermal optically thick disk emission. In XMMSL1 J0740-85, the 2-10 keV luminosity is $\sim 5 \times 10^{42}$ erg s$^{-1}$, which is about an order of magnitude lower than what would be required in Cygnus A-2. The TDE in Cygnus A-2 would therefore have to rather bright for a thermal TDE, and it might fall in what \cite{Auchettl2017} identified as the 'reprocessing valley': an observed luminosity gap between the thermal and jetted TDE events. For TDE's of such brightness, any X-rays are expected to be reprocessed into UV and optical wavelengths before escaping the source.

If Cygnus A-2 is a thermal TDE, we would expect there to be a thermal emission component below 2 keV, that is more luminous than the non-thermal emission component. {\it NuSTAR} does not observe at energies below 3 keV, leaving the 2013 {\it Swift} XRT observation as the only one that could have potentially seen 0.5-2.0 keV emission from the TDE. We therefore compared the {\it Swift} spectra from 2007 and 2013 between 0.5 and 2 keV. The 0.5-2.0 keV count rate in 2013 is roughly 13\% higher compared to 2007.

We first fit an absorbed thermal model to the 2007 spectrum, which yields a temperature of 3.1 keV and an abundance of 0.6 Z$_\odot$. We then fit the same thermal model plus an additional absorbed power law to the 2013 spectrum, fixing all of the parameters of the thermal model to the 2007 values. We also set an upper limit to the additional absorbing column of $10^{22}$ cm$^{-2}$. We find that an absorbed power law with $\Gamma = 1.5 $ and $N_H = 0.74 \times 10^{22}$ cm$^{-2}$ gives a good fit to the data, with a reduced $\chi^{2} = 19/21$. The intrinsic 0.5-2.0 keV luminosity of the powerlaw component is $1.8 \times 10^{43}$ erg s$^{-1}$. These parameters are all roughly consistent with a TDE spectrum. It is also important to note that the observed rise in 0.5-2.0 keV could not have been caused by the primary AGN, unless the absorbing column was significantly lower during those observations. However, from the concurrent {\it NuSTAR} observation in 2013 we know that this is not the case, further supporting the idea that the rise in X-ray luminosity in the {\it Swift} and {\it NuSTAR} spectra is connected to Cygnus A-2. 

Radio emission has been observed from only 6 TDE's so far. In all of the jetted TDE events, like the aforementioned Sw J1644+57 and Sw J2058+05, as well as Sw J1112.2 \citep{Brown2017}, 5GHz radio luminosities of $10^{40}-10^{42}$ erg s$^{-1}$ have been observed. In these events, the radio emission, like the X-ray emission, is thought to arise from the relativistic jets that are oriented toward our line of sight. In 3 other thermal TDE events, radio emission has been observed with luminosity ranges of $10^{37}-10^{39}$ erg s$^{-1}$. The origin of this emission is uncertain, although several models exist, such as transient jets \citep{Velzen2016} and non-relativistic winds \citep{Alexander2016}. The radio luminosity at 5 GHz of Cygnus A-2 is $\approx 6 \times 10^{39}$ erg s$^{-1}$. It is worth noting that the radio luminosity, like the X-ray luminosity, falls right in the middle of a bi-modal luminosity distribution, between the jetted and thermal TDE's. This suggests that if Cygnus A-2 is a TDE, it is unlike previously observed TDE's.

As we have shown in section \ref{AGN:subsec:wind}, we see no evidence in the {\it Chandra} data for the fast, ionized wind that {\it NuSTAR} detected in 2013. Similar types of outflow have been detected in the TDE's J1521+0749 \citep{Lin2015} and ASASSN-14li \citep{Kara2018}. They are interpreted as warm absorbers that are fast-moving, with speeds of $\sim 0.12c$ and $\sim 0.22c$ respectively. Additionally, both outflows were detected to be ionized, with $\log \xi \sim 2-3$ erg cm s$^{-1}$. This resembles the outflow in the \textit{NuSTAR} observation, with a velocity of $\sim 0.06c$ and $\log \xi \sim 3.2$. The {\it NuSTAR} data also suggests a second wind component with $v \sim 0.16c$, although this component is very poorly constrained. 

In ASASSN-14li, the outflow was not detected a year later, suggesting these outflows are naturally short-lived and stop once the accretion rate drops back below the Eddington rate. The timescale of approximately a year is consistent with the {\it Chandra} non-detection of the wind a few years after {\it NuSTAR}.  

For the outflow to exist in the first place, it must be super-Eddington. \cite{Reynolds2015} calculated the kinetic energy flux of the wind was to be $L_K > 1.7 \times 10^{45}$ erg s$^{-1}$. This is below the bolometric luminosity of the primary AGN. However, in a black hole a few orders of magnitudes less massive, this kinetic energy flux well be super-Eddington.

\section{Conclusion}
\label{AGN:sec:conc}

The absence of X-ray emission from Cygnus A-2 exceeding 5\% of the emission from the primary AGN in 2015-2017 \textit{Chandra} observations presents a new constraint on the origin of the transient. In particular, if Cygnus A-2 is a steadily accreting black hole, we would expect its X-ray flux to exceed this limit, unless it is either implausibly massive, $\sim 10^{8-9} M_{\odot}$, or if the absorbing column is significantly higher than the primary AGN at $2 \times 10^{23}$ cm$^{-2}$. In the latter case, the intrinsic luminosity could be a few times higher than the upper limit that we have set. 

The lack of X-ray radiation is more naturally explained if Cygnus A-2 is the radio afterglow of a destructive event. \cite{Perley2017} discuss the possibility of a supernova, although the high radio luminosity, the spatial correlation with a previously detected infrared point source, and the lack of variability over the year that the source was monitored with the VLA are all difficult to explain with supernova models. They therefore favor the explanation of a TDE in an offset, secondary AGN. A TDE scenario would not require the secondary black hole to be as massive as in a steadily accreting black hole scenario. Although TDE's are quite rare, 460 pc separation between the primary AGN and Cygnus A-2 might be enough to significantly enhance the average disruption rate of $10^{-4}$ - $10^{-5}$ yrs$^{-1}$ galaxy$^{-1}$.  

The X-ray light curve of the AGN of Cygnus A that we have constructed in this work, shows that the 2-10 keV luminosity has been fairly steady since at least 2000, with the exception of 2013, when the luminosity was twice as high. Although it is possible that the luminosity increase can be attributed to stochastic variability that is inherent to AGN, we suggest that Cygnus A-2 might instead have been responsible for this extra X-ray emission. We therefore investigated a TDE model for Cygnus A-2 based on the X-ray observations. To explain the 2-10 keV emission, the TDE would either have to be a powerful, jetted TDE or a thermal TDE with a hot corona comptonizing thermal disk emission and re-emitting it as non-thermal power law emission.

The observed 2-10 keV luminosity increase, of $0.5 - 1 \times 10^{44}$ erg s$^{-1}$, falls in the gap between typical luminosities for a thermal TDE with a non-thermal component ($10^{41}-10^{43}$ erg s${-1}$) and a jetted TDE ($10^{46}-10^{48}$ erg s$^{-1}$). Similarly, the radio luminosity, which at 5 GHz is $5 \times 10^{39}$ erg s$^{-1}$, falls in between the $10^{37}-10^{39}$ erg s$^{-1}$ luminosity range for thermal TDE's, and the $10^{40}-10^{42}$ erg s$^{-1}$ luminosity range for jetted TDE's. Cygnus A-2 would therefore either have to be a particularly powerful thermal TDE, or particularly faint jetted TDE, perhaps because of a larger viewer angle towards the jet. 

We have observed a small increase in 0.5-2.0 keV luminosity in the {\it Swift} XRT spectra between 2007 and 2013. This increase could not have been caused by the AGN itself, if the absorbing column is indeed as large as measured in 2013 by {\it NuSTAR}. Although the {\it Swift} data is not very constraining, it suggests a connection between the 2013 X-ray observations and Cygnus A-2. The detection of the fast, ionized wind in the 2013 {\it NuSTAR} data, and the non-detection with {\it Chandra} in 2015, can also be explained by a TDE, as these have been known to launch short-lived fast ionized winds. 

The sparsity of X-ray data around 2013, other than the {\it NuSTAR} and {\it Swift} observations discussed in this paper, make a more detailed X-ray analysis difficult. Additional observations between 2013 and 2015 could have confirmed whether the post-2013 light curve follows a TDE-like slope. Unequivocal proof that the X-ray emission in 2013 came from a TDE might therefore be out of reach. However, future X-ray observations of the AGN could potentially determine whether the X-ray luminosity rises again, or if perhaps the wind detected by {\it NuSTAR} has reappeared. If so, then this would imply that the X-ray luminosity peak and the fast wind, are recurring behaviour of the AGN rather than a unique event like a TDE.  Continued monitoring with the VLA and the VLBA should be able to confirm whether Cygnus A-2 is indeed a TDE.

\section*{Acknowledgements}

Support for this work was provided by the National Aeronautics and Space Administration through Chandra Award Number GO5-16117A issued by the Chandra X-ray Observatory Center, which is operated by the Smithsonian Astrophysical Observatory for and on behalf of the National Aeronautics Space Administration under contract NAS8-03060. This work made use of data supplied by the UK Swift Science Data Centre at the University of Leicester. This research has made use of data, software and/or web tools obtained from the High Energy Astrophysics Science Archive Research Center (HEASARC), a service of the Astrophysics Science Division at NASA/GSFC and of the Smithsonian Astrophysical Observatory's High Energy Astrophysics Division.



\bibliographystyle{mnras}
\bibliography{CygA_transient} 

\begin{thebibliography}{}
\makeatletter
\relax
\def\mn@urlcharsother{\let\do\@makeother \do\$\do\&\do\#\do\^\do\_\do\%\do\~}
\def\mn@doi{\begingroup\mn@urlcharsother \@ifnextchar [ {\mn@doi@}
  {\mn@doi@[]}}
\def\mn@doi@[#1]#2{\def\@tempa{#1}\ifx\@tempa\@empty \href
  {http://dx.doi.org/#2} {doi:#2}\else \href {http://dx.doi.org/#2} {#1}\fi
  \endgroup}
\def\mn@eprint#1#2{\mn@eprint@#1:#2::\@nil}
\def\mn@eprint@arXiv#1{\href {http://arxiv.org/abs/#1} {{\tt arXiv:#1}}}
\def\mn@eprint@dblp#1{\href {http://dblp.uni-trier.de/rec/bibtex/#1.xml}
  {dblp:#1}}
\def\mn@eprint@#1:#2:#3:#4\@nil{\def\@tempa {#1}\def\@tempb {#2}\def\@tempc
  {#3}\ifx \@tempc \@empty \let \@tempc \@tempb \let \@tempb \@tempa \fi \ifx
  \@tempb \@empty \def\@tempb {arXiv}\fi \@ifundefined
  {mn@eprint@\@tempb}{\@tempb:\@tempc}{\expandafter \expandafter \csname
  mn@eprint@\@tempb\endcsname \expandafter{\@tempc}}}

\bibitem[\protect\citeauthoryear{{Alexander}, {Berger}, {Guillochon},
  {Zauderer}  \& {Williams}}{{Alexander} et~al.}{2016}]{Alexander2016}
{Alexander} K.~D.,  {Berger} E.,  {Guillochon} J.,  {Zauderer} B.~A.,
  {Williams} P.~K.~G.,  2016, \mn@doi [\apjl] {10.3847/2041-8205/819/2/L25},
  \href {http://adsabs.harvard.edu/abs/2016ApJ...819L..25A} {819, L25}

\bibitem[\protect\citeauthoryear{{Antonucci}, {Hurt}  \& {Kinney}}{{Antonucci}
  et~al.}{1994}]{Antonucci1994}
{Antonucci} R.,  {Hurt} T.,   {Kinney} A.,  1994, \mn@doi [\nat]
  {10.1038/371313a0}, \href {http://adsabs.harvard.edu/abs/1994Natur.371..313A}
  {371, 313}

\bibitem[\protect\citeauthoryear{{Arnaud}, {Johnstone}, {Fabian}, {Crawford},
  {Nulsen}, {Shafer}  \& {Mushotzky}}{{Arnaud} et~al.}{1987}]{Arnaud1987}
{Arnaud} K.~A.,  {Johnstone} R.~M.,  {Fabian} A.~C.,  {Crawford} C.~S.,
  {Nulsen} P.~E.~J.,  {Shafer} R.~A.,   {Mushotzky} R.~F.,  1987, \mn@doi
  [\mnras] {10.1093/mnras/227.1.241}, \href
  {http://adsabs.harvard.edu/abs/1987MNRAS.227..241A} {227, 241}

\bibitem[\protect\citeauthoryear{{Auchettl}, {Guillochon}  \&
  {Ramirez-Ruiz}}{{Auchettl} et~al.}{2017}]{Auchettl2017}
{Auchettl} K.,  {Guillochon} J.,   {Ramirez-Ruiz} E.,  2017, \mn@doi [\apj]
  {10.3847/1538-4357/aa633b}, \href
  {http://adsabs.harvard.edu/abs/2017ApJ...838..149A} {838, 149}

\bibitem[\protect\citeauthoryear{{Bianchi}, {Piconcelli}, {Chiaberge},
  {Bail{\'o}n}, {Matt}  \& {Fiore}}{{Bianchi} et~al.}{2009}]{Bianchi2009}
{Bianchi} S.,  {Piconcelli} E.,  {Chiaberge} M.,  {Bail{\'o}n} E.~J.,  {Matt}
  G.,   {Fiore} F.,  2009, \mn@doi [\apj] {10.1088/0004-637X/695/1/781}, \href
  {http://adsabs.harvard.edu/abs/2009ApJ...695..781B} {695, 781}

\bibitem[\protect\citeauthoryear{{Bright} et~al.,}{{Bright}
  et~al.}{2018}]{Bright2018b}
{Bright} J.~S.,  et~al., 2018, \mnras, 475, 4011

\bibitem[\protect\citeauthoryear{{Brown} et~al.,}{{Brown}
  et~al.}{2017}]{Brown2017}
{Brown} G.~C.,  et~al., 2017, \mn@doi [\mnras] {10.1093/mnras/stx2193}, \href
  {http://adsabs.harvard.edu/abs/2017MNRAS.472.4469B} {472, 4469}

\bibitem[\protect\citeauthoryear{{Burrows} et~al.,}{{Burrows}
  et~al.}{2005}]{Burrows2005}
{Burrows} D.~N.,  et~al., 2005, \mn@doi [\ssr] {10.1007/s11214-005-5097-2},
  \href {http://adsabs.harvard.edu/abs/2005SSRv..120..165B} {120, 165}

\bibitem[\protect\citeauthoryear{{Burrows} et~al.,}{{Burrows}
  et~al.}{2011}]{Burrows2011}
{Burrows} D.~N.,  et~al., 2011, \mn@doi [\nat] {10.1038/nature10374}, \href
  {http://adsabs.harvard.edu/abs/2011Natur.476..421B} {476, 421}

\bibitem[\protect\citeauthoryear{{Canalizo}, {Max}, {Whysong}, {Antonucci}  \&
  {Dahm}}{{Canalizo} et~al.}{2003}]{Canalizo2003}
{Canalizo} G.,  {Max} C.,  {Whysong} D.,  {Antonucci} R.,   {Dahm} S.~E.,
  2003, \mn@doi [\apj] {10.1086/378513}, \href
  {http://adsabs.harvard.edu/abs/2003ApJ...597..823C} {597, 823}

\bibitem[\protect\citeauthoryear{{Davis}}{{Davis}}{2001}]{Davis2001}
{Davis} J.~E.,  2001, \mn@doi [\apj] {10.1086/323488}, \href
  {http://adsabs.harvard.edu/abs/2001ApJ...562..575D} {562, 575}

\bibitem[\protect\citeauthoryear{{Davis} et~al.,}{{Davis}
  et~al.}{2012}]{Davis2012}
{Davis} J.~E.,  et~al., 2012, in Space Telescopes and Instrumentation 2012:
  Ultraviolet to Gamma Ray. p. 84431A, \mn@doi{10.1117/12.926937}

\bibitem[\protect\citeauthoryear{{Dickey} \& {Lockman}}{{Dickey} \&
  {Lockman}}{1990}]{Dickey1990}
{Dickey} J.~M.,  {Lockman} F.~J.,  1990, \mn@doi [\araa]
  {10.1146/annurev.aa.28.090190.001243}, \href
  {http://adsabs.harvard.edu/abs/1990ARA%26A..28..215D} {28, 215}

\bibitem[\protect\citeauthoryear{{Franceschini}, {Vercellone}  \&
  {Fabian}}{{Franceschini} et~al.}{1998}]{Franceschini1998}
{Franceschini} A.,  {Vercellone} S.,   {Fabian} A.~C.,  1998, \mn@doi [\mnras]
  {10.1046/j.1365-8711.1998.01534.x}, \href
  {http://adsabs.harvard.edu/abs/1998MNRAS.297..817F} {297, 817}

\bibitem[\protect\citeauthoryear{{Freeman}, {Doe}  \&
  {Siemiginowska}}{{Freeman} et~al.}{2001}]{Freeman2001}
{Freeman} P.,  {Doe} S.,   {Siemiginowska} A.,  2001, in {Starck} J.-L.,
  {Murtagh} F.~D.,  eds,  \procspie Vol. 4477, Astronomical Data Analysis. pp
  76--87 (\mn@eprint {} {astro-ph/0108426}), \mn@doi{10.1117/12.447161}

\bibitem[\protect\citeauthoryear{{Fruscione} et~al.,}{{Fruscione}
  et~al.}{2006}]{Fruscione2006}
{Fruscione} A.,  et~al., 2006, in Society of Photo-Optical Instrumentation
  Engineers (SPIE) Conference Series. p. 62701V, \mn@doi{10.1117/12.671760}

\bibitem[\protect\citeauthoryear{{Gehrels} et~al.,}{{Gehrels}
  et~al.}{2004}]{Gehrels2004}
{Gehrels} N.,  et~al., 2004, \mn@doi [\apj] {10.1086/422091}, \href
  {http://adsabs.harvard.edu/abs/2004ApJ...611.1005G} {611, 1005}

\bibitem[\protect\citeauthoryear{{Gonz{\'a}lez-Mart{\'{\i}}n} \&
  {Vaughan}}{{Gonz{\'a}lez-Mart{\'{\i}}n} \& {Vaughan}}{2012}]{Gonzalez2012}
{Gonz{\'a}lez-Mart{\'{\i}}n} O.,  {Vaughan} S.,  2012, \mn@doi [\aap]
  {10.1051/0004-6361/201219008}, \href
  {http://adsabs.harvard.edu/abs/2012A%26A...544A..80G} {544, A80}

\bibitem[\protect\citeauthoryear{{Gordon} et~al.,}{{Gordon}
  et~al.}{2016}]{Gordon2016}
{Gordon} D.,  et~al., 2016, VizieR Online Data Catalog, \href
  {http://adsabs.harvard.edu/abs/2016yCat..51510154G} {515}

\bibitem[\protect\citeauthoryear{{Harrison} et~al.,}{{Harrison}
  et~al.}{2013}]{Harrison2013}
{Harrison} F.~A.,  et~al., 2013, \mn@doi [\apj] {10.1088/0004-637X/770/2/103},
  \href {http://adsabs.harvard.edu/abs/2013ApJ...770..103H} {770, 103}

\bibitem[\protect\citeauthoryear{{Hinshaw} et~al.,}{{Hinshaw}
  et~al.}{2013}]{Hinshaw2013}
{Hinshaw} G.,  et~al., 2013, \mn@doi [\apjs] {10.1088/0067-0049/208/2/19},
  \href {http://adsabs.harvard.edu/abs/2013ApJS..208...19H} {208, 19}

\bibitem[\protect\citeauthoryear{{Jackson}, {Tadhunter}  \& {Sparks}}{{Jackson}
  et~al.}{1998}]{Jackson1998}
{Jackson} N.,  {Tadhunter} C.,   {Sparks} W.~B.,  1998, \mn@doi [\mnras]
  {10.1046/j.1365-8711.1998.02008.x}, \href
  {http://adsabs.harvard.edu/abs/1998MNRAS.301..131J} {301, 131}

\bibitem[\protect\citeauthoryear{{Kara}, {Dai}, {Reynolds}  \&
  {Kallman}}{{Kara} et~al.}{2018}]{Kara2018}
{Kara} E.,  {Dai} L.,  {Reynolds} C.~S.,   {Kallman} T.,  2018, \mn@doi
  [\mnras] {10.1093/mnras/stx3004}, \href
  {http://adsabs.harvard.edu/abs/2018MNRAS.474.3593K} {474, 3593}

\bibitem[\protect\citeauthoryear{{Kawamuro}, {Ueda}, {Shidatsu}, {Hori},
  {Kawai}, {Negoro}  \& {Mihara}}{{Kawamuro} et~al.}{2016}]{Kawamuro2016}
{Kawamuro} T.,  {Ueda} Y.,  {Shidatsu} M.,  {Hori} T.,  {Kawai} N.,  {Negoro}
  H.,   {Mihara} T.,  2016, \mn@doi [\pasj] {10.1093/pasj/psw056}, \href
  {http://adsabs.harvard.edu/abs/2016PASJ...68...58K} {68, 58}

\bibitem[\protect\citeauthoryear{{Kelly}, {Sobolewska}  \&
  {Siemiginowska}}{{Kelly} et~al.}{2011}]{Kelly2011}
{Kelly} B.~C.,  {Sobolewska} M.,   {Siemiginowska} A.,  2011, \mn@doi [\apj]
  {10.1088/0004-637X/730/1/52}, \href
  {http://adsabs.harvard.edu/abs/2011ApJ...730...52K} {730, 52}

\bibitem[\protect\citeauthoryear{{Landt}, {Ward}, {Peterson}, {Bentz}, {Elvis},
  {Korista}  \& {Karovska}}{{Landt} et~al.}{2013}]{Landt2013}
{Landt} H.,  {Ward} M.~J.,  {Peterson} B.~M.,  {Bentz} M.~C.,  {Elvis} M.,
  {Korista} K.~T.,   {Karovska} M.,  2013, \mn@doi [\mnras]
  {10.1093/mnras/stt421}, \href
  {http://adsabs.harvard.edu/abs/2013MNRAS.432..113L} {432, 113}

\bibitem[\protect\citeauthoryear{{Lin} et~al.,}{{Lin} et~al.}{2015}]{Lin2015}
{Lin} D.,  et~al., 2015, \mn@doi [\apj] {10.1088/0004-637X/811/1/43}, \href
  {http://adsabs.harvard.edu/abs/2015ApJ...811...43L} {811, 43}

\bibitem[\protect\citeauthoryear{{Liu} \& {Chen}}{{Liu} \&
  {Chen}}{2013}]{Liu2013}
{Liu} F.~K.,  {Chen} X.,  2013, \mn@doi [\apj] {10.1088/0004-637X/767/1/18},
  \href {http://adsabs.harvard.edu/abs/2013ApJ...767...18L} {767, 18}

\bibitem[\protect\citeauthoryear{{Markowitz} et~al.,}{{Markowitz}
  et~al.}{2003}]{Markowitz2003}
{Markowitz} A.,  et~al., 2003, \mn@doi [\apj] {10.1086/375330}, \href
  {http://adsabs.harvard.edu/abs/2003ApJ...593...96M} {593, 96}

\bibitem[\protect\citeauthoryear{{Mushotzky}, {Done}  \& {Pounds}}{{Mushotzky}
  et~al.}{1993}]{Mushotzky1993}
{Mushotzky} R.~F.,  {Done} C.,   {Pounds} K.~A.,  1993, \mn@doi [\araa]
  {10.1146/annurev.astro.31.1.717}, \href
  {http://adsabs.harvard.edu/abs/1993ARA%26A..31..717M} {31, 717}

\bibitem[\protect\citeauthoryear{{Ogle}, {Cohen}, {Miller}, {Tran}, {Fosbury}
  \& {Goodrich}}{{Ogle} et~al.}{1997}]{Ogle1997}
{Ogle} P.~M.,  {Cohen} M.~H.,  {Miller} J.~S.,  {Tran} H.~D.,  {Fosbury}
  R.~A.~E.,   {Goodrich} R.~W.,  1997, \mn@doi [\apjl] {10.1086/310675}, \href
  {http://adsabs.harvard.edu/abs/1997ApJ...482L..37O} {482, L37}

\bibitem[\protect\citeauthoryear{{Pasham} et~al.,}{{Pasham}
  et~al.}{2015}]{Pasham2015}
{Pasham} D.~R.,  et~al., 2015, \mn@doi [\apj] {10.1088/0004-637X/805/1/68},
  \href {http://adsabs.harvard.edu/abs/2015ApJ...805...68P} {805, 68}

\bibitem[\protect\citeauthoryear{{Perley}, {Perley}, {Dhawan}  \&
  {Carilli}}{{Perley} et~al.}{2017}]{Perley2017}
{Perley} D.~A.,  {Perley} R.~A.,  {Dhawan} V.,   {Carilli} C.~L.,  2017,
  preprint, \href {http://adsabs.harvard.edu/abs/2017arXiv170507901P} {}
  (\mn@eprint {arXiv} {1705.07901})

\bibitem[\protect\citeauthoryear{{Plotkin}, {Markoff}, {Kelly}, {K{\"o}rding}
  \& {Anderson}}{{Plotkin} et~al.}{2012}]{Plotkin2012}
{Plotkin} R.~M.,  {Markoff} S.,  {Kelly} B.~C.,  {K{\"o}rding} E.,   {Anderson}
  S.~F.,  2012, \mn@doi [\mnras] {10.1111/j.1365-2966.2011.19689.x}, \href
  {http://adsabs.harvard.edu/abs/2012MNRAS.419..267P} {419, 267}

\bibitem[\protect\citeauthoryear{{Privon}, {Baum}, {O'Dea}, {Gallimore},
  {Noel-Storr}, {Axon}  \& {Robinson}}{{Privon} et~al.}{2012}]{Privon2012}
{Privon} G.~C.,  {Baum} S.~A.,  {O'Dea} C.~P.,  {Gallimore} J.,  {Noel-Storr}
  J.,  {Axon} D.~J.,   {Robinson} A.,  2012, \mn@doi [\apj]
  {10.1088/0004-637X/747/1/46}, \href
  {http://adsabs.harvard.edu/abs/2012ApJ...747...46P} {747, 46}

\bibitem[\protect\citeauthoryear{{Reynolds} et~al.,}{{Reynolds}
  et~al.}{2015}]{Reynolds2015}
{Reynolds} C.~S.,  et~al., 2015, \mn@doi [\apj] {10.1088/0004-637X/808/2/154},
  \href {http://adsabs.harvard.edu/abs/2015ApJ...808..154R} {808, 154}

\bibitem[\protect\citeauthoryear{{Sambruna}, {Eracleous}  \&
  {Mushotzky}}{{Sambruna} et~al.}{1999}]{Sambruna1999}
{Sambruna} R.~M.,  {Eracleous} M.,   {Mushotzky} R.~F.,  1999, \mn@doi [\apj]
  {10.1086/307981}, \href {http://adsabs.harvard.edu/abs/1999ApJ...526...60S}
  {526, 60}

\bibitem[\protect\citeauthoryear{{Saxton}, {Read}, {Komossa}, {Lira},
  {Alexander}  \& {Wieringa}}{{Saxton} et~al.}{2017}]{Saxton2017}
{Saxton} R.~D.,  {Read} A.~M.,  {Komossa} S.,  {Lira} P.,  {Alexander} K.~D.,
  {Wieringa} M.~H.,  2017, \mn@doi [\aap] {10.1051/0004-6361/201629015}, \href
  {http://adsabs.harvard.edu/abs/2017A%26A...598A..29S} {598, A29}

\bibitem[\protect\citeauthoryear{{Snios} et~al.,}{{Snios}
  et~al.}{2018}]{Snios2018}
{Snios} B.,  et~al., 2018, \mn@doi [\apj] {10.3847/1538-4357/aaaf1a}, \href
  {http://adsabs.harvard.edu/abs/2018ApJ...855...71S} {855, 71}

\bibitem[\protect\citeauthoryear{{Sobolewska} \& {Papadakis}}{{Sobolewska} \&
  {Papadakis}}{2009}]{Sobolewska2009}
{Sobolewska} M.~A.,  {Papadakis} I.~E.,  2009, \mn@doi [\mnras]
  {10.1111/j.1365-2966.2009.15382.x}, \href
  {http://adsabs.harvard.edu/abs/2009MNRAS.399.1597S} {399, 1597}

\bibitem[\protect\citeauthoryear{{Soldi} et~al.,}{{Soldi}
  et~al.}{2014}]{Soldi2014}
{Soldi} S.,  et~al., 2014, \mn@doi [\aap] {10.1051/0004-6361/201322653}, \href
  {http://adsabs.harvard.edu/abs/2014A%26A...563A..57S} {563, A57}

\bibitem[\protect\citeauthoryear{{Stockton}, {Ridgway}  \& {Lilly}}{{Stockton}
  et~al.}{1994}]{Stockton1994}
{Stockton} A.,  {Ridgway} S.~E.,   {Lilly} S.~J.,  1994, \mn@doi [\aj]
  {10.1086/117080}, \href {http://adsabs.harvard.edu/abs/1994AJ....108..414S}
  {108, 414}

\bibitem[\protect\citeauthoryear{{Tadhunter}, {Metz}  \&
  {Robinson}}{{Tadhunter} et~al.}{1994}]{Tadhunter1994}
{Tadhunter} C.~N.,  {Metz} S.,   {Robinson} A.,  1994, \mn@doi [\mnras]
  {10.1093/mnras/268.4.989}, \href
  {http://adsabs.harvard.edu/abs/1994MNRAS.268..989T} {268, 989}

\bibitem[\protect\citeauthoryear{{Tadhunter}, {Marconi}, {Axon}, {Wills},
  {Robinson}  \& {Jackson}}{{Tadhunter} et~al.}{2003}]{Tadhunter2003}
{Tadhunter} C.,  {Marconi} A.,  {Axon} D.,  {Wills} K.,  {Robinson} T.~G.,
  {Jackson} N.,  2003, \mn@doi [\mnras] {10.1046/j.1365-8711.2003.06588.x},
  \href {http://adsabs.harvard.edu/abs/2003MNRAS.342..861T} {342, 861}

\bibitem[\protect\citeauthoryear{{Tadhunter}, {Spence}, {Rose}, {Mullaney}  \&
  {Crowther}}{{Tadhunter} et~al.}{2017}]{Tadhunter2017}
{Tadhunter} C.,  {Spence} R.,  {Rose} M.,  {Mullaney} J.,   {Crowther} P.,
  2017, \mn@doi [Nature Astronomy] {10.1038/s41550-017-0061}, \href
  {http://adsabs.harvard.edu/abs/2017NatAs...1E..61T} {1, 0061}

\bibitem[\protect\citeauthoryear{{Turner} et~al.,}{{Turner}
  et~al.}{1989}]{Turner1989}
{Turner} M.~J.~L.,  et~al., 1989, \pasj, \href
  {http://adsabs.harvard.edu/abs/1989PASJ...41..345T} {41, 345}

\bibitem[\protect\citeauthoryear{{Ueno}, {Koyama}, {Nishida}, {Yamauchi}  \&
  {Ward}}{{Ueno} et~al.}{1994}]{Ueno1994}
{Ueno} S.,  {Koyama} K.,  {Nishida} M.,  {Yamauchi} S.,   {Ward} M.~J.,  1994,
  \mn@doi [\apjl] {10.1086/187458}, \href
  {http://adsabs.harvard.edu/abs/1994ApJ...431L...1U} {431, L1}

\bibitem[\protect\citeauthoryear{{Young}, {Wilson}, {Terashima}, {Arnaud}  \&
  {Smith}}{{Young} et~al.}{2002}]{Young2002}
{Young} A.~J.,  {Wilson} A.~S.,  {Terashima} Y.,  {Arnaud} K.~A.,   {Smith}
  D.~A.,  2002, \mn@doi [\apj] {10.1086/324200}, \href
  {http://adsabs.harvard.edu/abs/2002ApJ...564..176Y} {564, 176}

\bibitem[\protect\citeauthoryear{{Zauderer}, {Berger}, {Margutti}, {Pooley},
  {Sari}, {Soderberg}, {Brunthaler}  \& {Bietenholz}}{{Zauderer}
  et~al.}{2013}]{Zauderer2013}
{Zauderer} B.~A.,  {Berger} E.,  {Margutti} R.,  {Pooley} G.~G.,  {Sari} R.,
  {Soderberg} A.~M.,  {Brunthaler} A.,   {Bietenholz} M.~F.,  2013, \mn@doi
  [\apj] {10.1088/0004-637X/767/2/152}, \href
  {http://adsabs.harvard.edu/abs/2013ApJ...767..152Z} {767, 152}

\bibitem[\protect\citeauthoryear{{van Velzen} et~al.,}{{van Velzen}
  et~al.}{2016}]{Velzen2016}
{van Velzen} S.,  et~al., 2016, \mn@doi [Science] {10.1126/science.aad1182},
  \href {http://adsabs.harvard.edu/abs/2016Sci...351...62V} {351, 62}

\makeatother
\end{thebibliography}

\bsp	
\label{lastpage}
\end{document}